\documentclass[letter,11pt]{article}
\pdfoutput=1 

\usepackage{jheppub} 
                     
\usepackage{siunitx} 

\usepackage{multirow} 
\usepackage{booktabs} 
\usepackage{longtable} 

\usepackage{graphicx}
\usepackage{subcaption}

\title{\boldmath Constructing the theory at the boundary, its dynamics and degrees of freedom}

\author[a,b]{Irais Rubalcava-Garcia}

\affiliation[a]{Physique Th\'eorique et Math\'ematique,
	 Universit\'e Libre de Bruxelles and International Solvay Institutes, Campus Plaine C.P. 231, B-1050 Bruxelles, Belgium}
\affiliation[b,1]{Departamento de Matem\'aticas, Instituto de Ciencias, Benem\'erita Universidad Aut\'onoma de Puebla, 72570 Puebla, Pue., M\'exico\note{Present address}}

\emailAdd{irais@fcfm.buap.mx}
\emailAdd{irais11@gmail.com}

\abstract{Given a theory in a region of space-time with boundaries, defined by a Lagrangian, we propose a method to find the corresponding theory at the boundary (either space-like or time-like), that allows us to identify the degrees of freedom at the boundary without any gauge fixing nor further assumptions a priori. We start by finding the corresponding action principle at the boundary. From this, we can find the corresponding equations of motion, the symmetries and degrees of freedom at the boundary. Then we characterise the degrees of freedom at the boundary.
We exemplify this method by analysing the broadly studied 3-dimensional Abelian Chern-Simons theory. We found, through the Hamiltonian framework, that the boundary theory has one degree of freedom that can be characterised by the chiral boson: the Quantum Hall edge modes. We also find a correspondence between a gauge fixing in the bulk with a boundary condition and discuss whether this is enough to have a well-posed action principle and the possible scenarios.\\

}

\begin{document} 
\maketitle
\flushbottom

\section{Introduction}

Recently the importance of the boundary symmetries of a gauge theory\footnote{Assuming that this gauge theory is well-defined in a space-time with boundary.} has been unveiled in a wide variety of systems from the quest of a quantum theory of gravity to the study of topological insulators in condensed matter systems. Also, the classical geometrical description of gauge theories in space-times with boundaries is a beautiful one that deserves attention on its own. Therefore the primary goal of the present work, to propose a method to find the corresponding theory at the boundary from which we can study its symmetries, dynamics, degrees of freedom, among other things.

If we want to study phenomena far away, as the ones recently detected by gravitational waves, we can think of ourselves as asymptotic observers. We can see this asymptotic region as an outer boundary of space-time, and \cite{Brown-Henneaux-1986, Corichi-Rubalcava-Vukasinac-Review, Troessaert-2013} have pointed out that we need to take extra care to study space-times with boundaries. In the case when dealing with asymptotically flat space-times it is well known that the symmetry group is larger than the Poincar\'e group (the symmetry group of Minkowski space-time). At null infinity the symmetry group is the semi-direct product of the group of supertranslations with the Lorentz group, that is called the BMS group named after Bondi, Metzner and Sachs \cite{Ashtekar-2015, Barnich-Troessaert-2010}.

The quest of a quantum theory for the gravitational field is still one of the biggest goals in theoretical physics. Also, in this context, the understanding of the symmetries at the boundaries play an important role. The structure of null infinity provides a natural arena for the S-matrix theory. In a recent article \cite{Temple-Lysov-Mitra-Strominger-2015}, the authors conjectured that a certain infinite-dimensional subgroup of the BMS group (acting on past and future null infinity) is an exact symmetry of the S-matrix for quantum gravity. They showed that the Ward identity associated with that symmetry is equivalent to Weinberg's soft graviton theorem. 
Moreover, these soft gravitons turn out to be the boundary modes, and they are manifestly the Goldstone bosons associated with such symmetry. Anticipating a bit the results of this work, we show that at the classical level, this is the general behaviour of the boundary dynamics, the boundary modes emerge due to a broken gauge symmetry in the bulk that is equivalent to a boundary condition. We also need the null infinity structure in the formulation of conceptual issues such as information loss during black hole evaporation and the role of the CPT symmetry in quantum gravity \cite{Ashtekar-Pretorius-Ramazanoglu-2011, Ashtekar-Taveras-Varadarajan-2008}.

The story does not end there, in the last few years a great deal of motivation to understand the symmetries at the boundary of field theories has come from Condensed Matter systems that can either be studied using field-theoretical methods or the Gauge-Gravity conjecture.  In particular, the latter has become very fruitful to solve problems in the strong-coupling regime, that otherwise were very difficult or impossible to address with the standard methods \cite{Maldacena1999, Papantonopoulos-2011}. One of the most popular versions of this duality conjecture is the AdS/CFT conjecture.

It has long been a puzzle why and how in some topological bulk theories (that have no local degrees of freedom in the bulk) the degrees of freedom emerge at the boundary. In this direction, there are recent works showing that the Chern-Simons has one degree of freedom at the boundary. For instance in \cite{Gallardo-Montesinos-2011} the authors show in a constructive way that when restricting ourselves to the space of solutions, the Chern-Simons action can be written as a total derivative (assuming that the topology of space-time allow us to do so), and then applying the Dirac algorithm to find the constraints and number of degrees of freedom. In \cite{Troessaert-2013} using covariant tools the authors are able to deduce the number of degrees of freedom at the boundary, however, their technique is not generalizable to theories with local degrees of freedom like gravity in 4 dimensions. The problem comes from the necessity of imposing boundary conditions a priori on the dynamical variables.

As seen from previous works, the standard way to approach these problems is to \emph{propose} boundary conditions a priori (inspired on the desired symmetries and on physical assumptions of the phenomena we want to study), reduce the theory and then extract the physical information. But in this way of proceeding, how do we know that we are not losing the more general setting and within it, interesting solutions? Supposing we have checked that our action principle is well defined under the imposed (by hand) boundary conditions. Other more subtle and technical aspects may arise: How we can be sure that the imposed boundary conditions do not propagate degrees of freedom outside the boundary? Within the standard approach, we are only finding all the boundary symmetries allowed by the imposed boundary conditions, but is there a way, given a theory defined by an action principle on a space-time with boundaries, to be able to know all the boundary symmetries and all the allowed boundary conditions without imposing conditions by hand, just the minimum required for the theory to be well posed?

At this point, it is important to emphasize that there is no such general method available at the moment in the literature. Most of the proposals in this line of thought have been applied for Chern-Simons theories, as the present work, because of its simplicity and the fact that we know so many things about it. That is the main reason to use this example, to be able to compare with other works and show the usefulness of the method we propose in this work.

The main goal of this work is, given a field theory defined in a space-time with boundaries, to construct its corresponding theory at the boundary. In particular, in what follows we shall focus in time-like boundaries (but this method can be applied to space-like boundaries as well). 

The idea is to project the space-time action principle to a time-like hypersurface, \emph{the boundary}. For this we need a minimum of assumptions, just the type of surface we are projecting on and its normal vector at each point. This can also be extended to a family of time-like hypersurfaces as needed for asymptotically flat spacetimes. We consider this latter approach here, since considering only one surface from the family is a particular case.

After the projection we need to ensure that the dynamics stays on the boundary, so we can have a well posed action that describes only the dynamics at the boundary. The resulting action principle at the boundary will provide a description of the dynamics and the symmetries at the boundary, as well as its degrees of freedom. 

The approach followed here applies to any theory with a local finite (or asymptotic) boundary. Among other things, we do not need to write the action as a total derivative (that would require certain topological conditions on the manifold on which is defined).

We find the allowed boundary conditions that make the action principle well-defined, and we show that the boundary conditions can be classified into two types. One that restricts the dynamics to the boundary, that we find systematically by solving  \emph{the condition of the restriction to the boundary} that we can read directly from the action, and the other that is needed to make the action principle well posed. The first one only depends on the action principle chosen and demanding that we do not want the propagation of the degrees of freedom from one time-like boundary to the next one. The second one also depends on the topology of the boundary and the solutions to the equations of motion we want to allow. Not all the possible solutions to the equations of motion are compatible with all the allowed boundary conditions. We need both types of boundary conditions to completely have a well posed action principle both in the bulk and in the boundary.

This article is organized as follows: 

In section \ref{Section:Constructing the theory at the boundary}, we present the proposed method to construct the boundary theory, given the bulk theory and the geometry of the boundary. We exemplify this method by studying the three-dimensional abelian Chern-Simons action, although our method is more general and can be applied to wide range of theories and boundaries, topological or with local degrees of freedom.

In section \ref{Section-Analyzing the theory at the boundary}, once that we have found the action principle at the boundary, we extract the relevant information, we analyze the well-posedness of the bulk and boundary action, and find the allowed boundary conditions that make them well-defined. We also study the dynamics at the boundary, i.e. the solutions to the equations of motion of the boundary action principle and the Hamiltonian analysis, where we find the constraints, degrees of freedom and gauge symmetries of both the bulk and boundary theories and we make a comparison. Finally, we discuss how we can recover in a direct way the known results for the Quantum Hall effect.

To make the paper more readable we left many details to the appendices.

In appendix \ref{Section-Geometrical-Preliminaries}, we introduce the necessary tools to make the projection of the space-time action to the boundary.  

In appendix \ref{Appendix-Dirac-bulk-theory}, we show the Hamiltonian analysis of the bulk action, although these results are known \cite{Escalante-Carbajal-2011}, we include them and express them in a way that can easily compare bulk with the boundary theory.

In appendix \ref{Hamiltonian-analysis-complete-theory} we make the Hamiltonian analysis of the theory at the boundary. We find the constraints, the degrees of freedom, the gauge symmetries. 

In appendix \ref{Appendix-Projecting-Bulk-EOM} we show that if, for some reason, we are only interested in the dynamics at the boundary given the bulk equations of motion we can find the boundary ones, and they agree with the ones derived from the boundary action principle.

We hope the main ideas of this article can be followed by non-experts in differential geometry nor general relativity, so they can be applied to a wider range of problems, specially in Condensed Matter systems.

\section{Constructing the theory at the boundary}\label{Section:Constructing the theory at the boundary}

Given a field theory defined in a space-time with boundaries, we want to construct, in a general and geometrical way, the corresponding theory at the boundary. With such a theory at the boundary, we shall be able to find the dynamics, the degrees of freedom, its symmetries without imposing particular boundary conditions a priori. Moreover, we shall be able to find the allowed boundary conditions compatible with the action principle.

Although the method proposed in the present article is as general as possible, we have to make a minimum of choices: To start, we need to choose the action principle in a region of a space-time with boundaries, and on which type of boundary we want to project this theory (time-like, space-like, null). These are the minimum requirements to construct the corresponding theory at the boundary. In brief, our proposed method goes as follows:
\begin{enumerate}
	\item {\textbf{Choosing an action principle:}} We chose an action principle for the desired theory defined on a region of a space-time with boundaries. For such action, we assume that it is well-posed under certain boundary conditions. Furthermore, as a byproduct, we will be able to identify those boundary conditions. 
	\item \textbf{Choosing the type of boundary:} The second key ingredient is to choose the type of boundary of interest. It is enough to specify if the boundary is time-like, space-like or null. We shall not discuss the null case in detail in this work. 
	\item \textbf{Projecting the bulk action principle:} Once we have chosen the action and the type of boundary, we need to project the action to the desired boundaries (hypersurfaces). The null case will be discussed elsewhere since there is not a unique way to project into a null surface, and extra care is needed.
	\item \textbf{Constructing the corresponding action principle at the boundary:} In order to have a well-defined action principle on the boundary, we need to ensure that there are no degrees of freedom propagating outside the hypersurface. This condition can be read directly from the projected action principle, and we need to impose it as a constraint. In this way, the initial value problem on the boundary will be well-posed. So we can study the dynamics on the boundary only.
	\item \textbf{Finding the relevant information at the boundary:} After we have constructed the corresponding theory at the boundary, given by an action principle, we can analyze it and get the relevant information. For instance, we can find its equations of motions and therefore, its dynamics at the boundary; or we can make a lagrangian analysis or a hamiltonian analysis. 
\end{enumerate}
We shall explain each point in more detail below. We exemplify this method with the three-dimensional abelian Chern-Simons action, but it can be applied to any field theory in any dimension defined on a region of a space-time with boundaries.

\subsection{Choosing the action principle}

Given a field theory defined on a space-time with boundaries, we know that there is not a unique choice of an action principle that leads to the same classical equations of motion in the bulk. In particular, two action principles that differ on a boundary term will be  classically equivalent in the bulk \cite{Rubalcava-Juarez-2019a}. Also, we can formulate a given theory with a different choice of variables that may result in different action principles not necessarily related by a boundary term. Therefore the first choice is to pick, among all the possibilities, one action principle to find its corresponding theory at the boundary.

Once we have chosen the action principle, we shall assume that it is well-posed under certain boundary conditions. In particular, we need it to be differentiable so that we can recover the Euler-Lagrange equations of motion. As we shall see below, as a by-product of this construction, we will be able to find boundary conditions compatible with the action principle, so we do not need to impose them a priori by hand.

However, once we choose an action principle, not all the boundary conditions are allowed (such that the action principle is well posed and that the dynamics at the boundary is constrained to it, and there is no ``leaking of degrees of freedom'' to one boundary to other). If we just imposed any ``boundary condition'' it may happen that we end up with an infinite number of constraints (See e.g. \cite{Romero-Vergara-2002,Sheikh-Jabbari-Shirzad-2001}).

As we mentioned in the introduction, we shall exemplify this method with the well known 3-dimensional Abelian Chern-Simons action, from which many relevant results are known using different approaches. So, in the end, we shall be able to compare our method and results with other approaches. We want to remark that this method is general and can be applied to a wide range of theories, with or without local degrees of freedom, in any dimension defined on a region of a space-time with boundaries.

\subsubsection{3-dimensional Abelian Chern-Simons action}

As already mentioned, we chose the well known three-dimensional abelian Chern-Simons action, \cite{Dunne-1999,Elitzur-etal-1989,Zanelli-2008-UsesCS,Zanelli-2012-LecturesCS},
\begin{equation}\label{CSAction-forms}
S_{CS} [A] = \frac{\kappa}{2} \int_{\mathcal{M}} \left( A \wedge \mathrm{d}A \right),
\end{equation}
or expressed in components,
\begin{equation}\label{CSAction-components}
S_{CS} [A] = \frac{\kappa}{2} \int_{\mathcal{M}} A_{\mu} \partial_{\nu} A_{\rho} \tilde{\varepsilon}^{\mu \nu \rho} d^3 x =  \frac{\kappa}{4} \int_{\mathcal{M}} A_{\mu} F_{\nu \rho} \tilde{\varepsilon}^{\mu \nu \rho} d^3 x,
\end{equation}
where $F_{\mu \nu} = \partial_{\mu} A_{\nu} - \partial_{\nu} A_{\mu} $ is the curvature of the connection $A$ valued on the Lie Algebra of $U(1)$, $\mu,\nu = 0,1,2$ are space-time indices and $\kappa$ is the Chern-Simons coupling parameter\footnote{In the case when matter is coupled to the Chern-Simons action this coupling parameter $\kappa$ is the proportionality constant between the \emph{magnetic field} and the \emph{charge density}. This in analogy with the definitions of three dimensional electrodynamics \cite{Dunne-1999}. We must assume that, $\kappa = \displaystyle \frac{integer}{4 \pi}$, if we want to ensure that the quantum amplitude $\exp (\mathrm{i}S)$ remains gauge invariant.}. With $\tilde{\varepsilon}^{\mu \nu \rho} $ the Levi-Civita tensor density of weight $-1$ (see for instance \cite{Carrol-2004,Poisson-2004}\footnote{Note the difference in sign with other references, especially within the Loop Quantum Gravity community whose convention is that the Levi-Civita tensor density $\tilde{\varepsilon}^{\mu \nu \rho} $ has weight $+1$ \cite{Bojowald-2010,Corichi-Rubalcava-2015}. Here we are following the standard convention used in some General Relativity textbooks \cite{Carrol-2004,Poisson-2004}.}), that is related with $\mathrm{d} x^{\mu} \wedge \mathrm{d} x^{\nu} \wedge \mathrm{d} x^{\rho}$ as follows, $\mathrm{d} x^{\mu} \wedge \mathrm{d} x^{\nu} \wedge \mathrm{d} x^{\rho} = \tilde{\varepsilon}^{\mu \nu \rho} \mathrm{d} ^{3} x$.  It is also related with the Levi-Civita tensor $\varepsilon^{\mu \nu \rho}$, $\tilde{\varepsilon}^{\mu \nu \rho} = (s) \sqrt{|g|} \varepsilon^{\mu \nu \rho}$, with $s$ the signature of the metric, when there is one\footnote{Although the Chern-Simons theory (defined on an odd dimension manifold) is a topological theory and therefore does not need a metric to be well-defined on a manifold with boundary. If we want to consider a time-like, space-like or null boundary, we need to introduce a metric. It is only when we want to project the theory on a particular boundary that we need to introduce a metric.}. We have to remark that in the literature it is standard to drop $\mathrm{d} ^{3} x$ in the expressions for simplicity in the notation and we follow that convention here.

\subsection{Choosing the type of boundary}

One of the advantages of this method is that it is local, that is, we do not need the boundary to be globally defined, it is enough that locally we can embed a hypersurface (that will act as our boundary). So we can always use this method in a region of the space-time where we can embed such hypersurface, even if it is not well defined in all the space-time. More over, we do not need the action to be written as a total derivative or boundary term to analyse the theory at the boundary.

It is important to emphasize that all the discussion here corresponds to physical boundaries. In a conductor can be the actual edges of the material. In an asymptotically flat space-time can be the asymptotic region. The case of virtual boundaries that split the space-time region $\mathcal{ M}$ into two or more regions with boundary is left for forthcoming discussions. 
\begin{figure}[ht]
	\begin{center}
		\includegraphics[width=4cm]{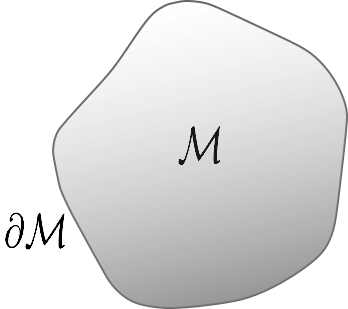}
		\caption{A region of space-time $\mathcal{ M}$ with its boundary $\partial \mathcal{ M}$.}\label{BoundaryHallState1}
	\end{center}
\end{figure}

We consider a region of a space-time, $\mathcal{M}$, such that it has topology $\mathbb{R}\times \mathcal{B}$, where $\mathcal{B}$ can be a space-like or time-like hypersurface. This means that $\mathcal{M}$ can be foliated by a family of hypersurfaces $(\mathcal{B}_{R})_{R\in \mathbb{R}}$ such that $\mathcal{ M} = \bigcup_{R \in \mathbb{R}} \mathcal{ B}_{R}.$ (see Appendix \ref{Section-Geometrical-Preliminaries} for all the mathematical definitions and details). 

\subsubsection{Time-like boundary}

This general method to construct the theory at the boundary is applicable to any hypersurface, however we shall restrict the present work to the cases for space-like and time-like boundaries. In particular, if we choose a time-like boundary we shall be able to have the standard ``time evolution'' and applied the well known methods, as the Hamiltonian analysis used here. 

In the example chosen to illustrate this method, the three-dimensional abelian Chern-Simons, there is no difference in our analysis between a space-like or time-like surface, this is because the theory per se does not need a metric to be defined, so there is no a notion of time a priori. But in other theories where a metric is needed to be defined, we shall be able to differentiate both cases.

\subsection{Projecting the bulk action principle}

The projected Levi-Civita tensor density is $\tilde{\varepsilon}^{\mu \nu \rho} = 3R^{[\mu}\tilde{\varepsilon}^{ \nu \rho]} \mathrm{d} R$. Note that this expression for the Levi-Civita tensor density holds for either a time-like or space-like surface. Inserting this projected Levi-Civita tensor density in the action (\ref{CSAction-components}) is equivalent to project the connection $A$ to the boundary. Then, the projected Chern-Simons action is,
\begin{eqnarray}\label{CSAction-projected0}
S_{CS} [A] &=&  \frac{\kappa}{4} \int_{\mathcal{M}} A_{\mu} F_{\nu \rho} 3 R^{[\mu}\tilde{\varepsilon}^{ \nu \rho]} \mathrm{d} R
\\	&=& \frac{\kappa}{4} \int_{\mathcal{M}} A_{\mu} F_{\nu \rho} \left( R^{\mu}\tilde{\varepsilon}^{ \nu \rho} + R^{\rho}\tilde{\varepsilon}^{ \mu \nu } + R^{\nu}\tilde{\varepsilon}^{ \rho \mu} \right)  \mathrm{d} R\\
&=& \frac{\kappa}{4} \int_{\mathcal{M}}  \left[  \left(R^{\mu} A_{\mu}\right) F_{\nu \rho}\tilde{\varepsilon}^{ \nu \rho} + 2 A_{\mu} F_{\nu \rho} R^{\rho}\tilde{\varepsilon}^{ \mu \nu } \right]  \mathrm{d} R
\end{eqnarray} 
using that\footnote{This expression is obtained using the identity $\pounds_{\vec{R}} A = i_{\vec{R}}\mathrm{d}A + \mathrm{d}(i_{\vec{R}}A)$ for differential forms, and the definition of $F=\mathrm{d} A$. It is usually written as $R^{\rho} F_{ \rho \nu } = \pounds_{\vec{R}} A_{\nu} - \nabla_{\nu} (R^{\rho} A_{\rho})  $.} $R^{\rho} F_{\nu \rho } = \partial_{\nu} (R^{\rho} A_{\rho}) - \pounds_{\vec{R}} A_{\nu} $, the previous equation becomes\footnote{Note that this action is a bulk action, completely equivalent to the original one, but now it also has the information of the time-like family of hypersurfaces we are projetcting on. Since the Chern-Simons theory is topological in the sense we do not need a metric to define it, and therefore there is no notion a priorio of space-like or time-like, we could use the previous expression to make a hamiltonian analysis taking the directon of evolution as the direction given by $\nabla R$ (see appendix \ref{Section-Geometrical-Preliminaries}).},
\begin{equation}\label{CS-Projected-action}
S_{CS} [A] =  \frac{\kappa}{4} \int_{\mathcal{M}} \left[ (A_{\rho} R^{\rho}) F_{\mu \nu} + 2 A_{\mu} \partial_{\nu} (A_{\rho} R^{\rho}) - 2 A_{\mu} \pounds_{\vec{R}} A_{\nu} \right] \tilde{\varepsilon}^{ \mu \nu} \mathrm{d} R.
\end{equation}

\subsection{Constructing the corresponding action principle at the boundary}

The last term in equation (\ref{CS-Projected-action}), $ 2 A_{\mu} \pounds_{\vec{R}} A_{\nu} \tilde{\varepsilon}^{ \mu \nu} $, contains the information about the propagation of the degrees of freedom along R, or in other words from one surface $\mathcal{B}_{R}$ to the next $\mathcal{B}_{R+\delta R}$. If we want to restrict the previous action to a particular surface $\mathcal{B}_{R}$, we need to get rid of the term that encodes the propagation of the degrees of freedom across the the surfaces $\mathcal{B}_{R}$. The naive way of doing so would be to drop the corresponding term from the action, but in that way we also lose some dynamical information on the variables. So, we need to use the Lagrange multiplier method and consider that term as a constraint to be fullfilled in each boundary $\mathcal{B}_{R}$. Taking into account this disscussion we can define the \emph{Chern-Simons action at the family of boundaries} $\mathcal{B}_{R}$ as,
\begin{eqnarray}\label{BoundaryAction-munu}
\nonumber S_{CS-B1} [A, \lambda] &=&  \frac{\kappa}{4} \int_{\mathcal{M}} \left[ (A_{\rho} R^{\rho}) F_{\mu \nu} + 2 A_{\mu} \partial_{\nu} (A_{\rho} R^{\rho}) - 2 A_{\mu} \pounds_{\vec{R}} A_{\nu} - \lambda\left( 2 A_{\mu} \pounds_{\vec{R}} A_{\nu} \right) \right] \tilde{\varepsilon}^{ \mu \nu} \mathrm{d} R \\
&=& \frac{\kappa}{4} \int_{\mathcal{M}} \left[ (A_{\rho} R^{\rho}) F_{\mu \nu} + 2 A_{\mu} \partial_{\nu} (A_{\rho} R^{\rho}) - 2 (1 + \lambda) A_{\mu} \pounds_{\vec{R}} A_{\nu}   \right] \tilde{\varepsilon}^{ \mu \nu} \mathrm{d} R.
\end{eqnarray}
It is important to remark that this is still a bulk action (that is why we are considering $\mathcal{ M}$ as the integration region). It contains information of all the slices, though they are independent among them (there are no propagating degrees of freedom from one time-like slice $\mathcal{B}_{R}$ to the next one $\mathcal{B}_{R+\delta R}$ since we have imposed \emph{the condition of the restriction to the boundary}, $- 2 \left( A_{i} \partial_{R} A_{j} \right)  \tilde{\varepsilon}^{ij} =0$).

Using coordinates adapted to the foliation, $\partial_{R} A_{j} = (\pounds_{\vec{R}} A_{j})$, now indices $i,j=0,1$ label arbitrary coordinates on each of the timelike boundaries, $\mathcal{B}_{R}$,
\begin{equation}\label{Chern-Simons-Boundary}
S_{CSB} [A_{R}, A_{i}, \lambda] =  \frac{\kappa}{4} \int_{\mathcal{M}} \left[ A_{R}  F_{ij} + 2 A_{i} \partial_{j} A_{R} - 2 (1 + \lambda) A_{i} \left( \partial_{R} A_{j}\right)  \right]  \tilde{\varepsilon}^{ij} \mathrm{d} R.
\end{equation}
This is our action that contains the information of the time-like hypersurfaces, each being independent from one another. After the corresponding analysis we can chose one of those hypersurfaces (or the limit if we are studying asymptotics) and recover the desired dynamics at the boundary.

\section{Analyzing the theory at the boundary}\label{Section-Analyzing the theory at the boundary}

In the previous section we have constructed \emph{the theory at the boundary}: 
\begin{equation*}
S_{CS-B1} [A, \lambda] = \frac{\kappa}{4} \int_{\mathcal{M}} \left[ (A_{\rho} R^{\rho}) F_{\mu \nu} + 2 A_{\mu} \partial_{\nu} (A_{\rho} R^{\rho}) - 2 (1 + \lambda) A_{\mu} \pounds_{\vec{R}} A_{\nu}   \right] \tilde{\varepsilon}^{ \mu \nu} \mathrm{d} R,
\end{equation*}
or in adapted coordinates to the foliation,
\begin{equation*}
S_{CSB} [A_{R}, A_{i}, \lambda] =  \frac{\kappa}{4} \int_{\mathcal{M}} \left[ A_{R}  F_{ij} + 2 A_{i} \partial_{j} A_{R} - 2 (1 + \lambda) A_{i} \left( \partial_{R} A_{j}\right)  \right]  \tilde{\varepsilon}^{ij} \mathrm{d} R.
\end{equation*}
With this action principle we shall be able to analyse it with different tools, depending what we want to study and our particular choice. For instance, we can only study the dynamics at the boundary, given by the equations of motion, and the compatible boundary conditions that make the action principle well posed. We can also make a lagrangian \cite{Diaz-Montesinos-2018} or hamiltonian anlysis \cite{Henneaux-Troessaert-2018ADM,Corichi-Rubalcava-Vukasinac-Review}. In this work we shall use the Dirac analysis to find the constraints, the gauge symmetries, the number of degrees of freedom and characterize them. 

\subsection{Well-posedness of the boundary action}\label{Subsection-Well-posedness-action-Boundary-conditions}

Using coordinates adapted to the foliation, $\partial_{R} A_{j} = (\pounds_{\vec{R}} A_{j})$, now indices $i,j=0,1$ label arbitrary coordinates on each of the timelike boundaries, $\mathcal{B}_{R}$,
\begin{equation}\label{Chern-Simons-Boundary}
S_{CSB} [A_{R}, A_{i}, \lambda] =  \frac{\kappa}{4} \int_{\mathcal{M}} \left[ A_{R}  F_{ij} + 2 A_{i} \partial_{j} A_{R} - 2 (1 + \lambda) A_{i} \left( \partial_{R} A_{j}\right)  \right]  \tilde{\varepsilon}^{ij} \mathrm{d} R.
\end{equation}
Taking the variation of the previous action,
\begin{eqnarray}\label{Variation SCB}
\nonumber &&\delta S_{CSB} [A_{R}, A_{i}, \lambda] =  \frac{\kappa}{2} \int_{\mathcal{M}} \left[ \partial_{i} \left( A_{R} \delta A_{j}- A_{j} \delta A_{R}  \right) - \frac{1}{2} \partial_{R} \left( (1 + \lambda)( A_{i} \delta A_{j} -  A_{j} \delta A_{i}) \right) \right. \\
\nonumber &&+  \left. \left( F_{ij} \right) \delta A_{R} - 2 \left(\partial_{i} A_{R} - \left(1 + \frac{\lambda}{2}\right) \partial_{R} A_{i} - \frac{1}{2} A_{i} \partial_{R} \lambda \right) \delta A_{j} - 2 \left( A_{i} \partial_{R} A_{j} \right) \delta \lambda \right] \tilde{\varepsilon}^{ij} \mathrm{d} R,
\end{eqnarray}
we can obtain the following equations of motion,
\begin{eqnarray}
\label{CSB-EqsMotion-F-0} \delta A_{R} &:& \tilde{\varepsilon}^{ij} F_{ij} = 0,\\
\label{CSB-EqsMotion-Gauss} \delta A_{j} &:&   - 2  \left(\partial_{i} A_{R} - \left(1 + \frac{\lambda}{2}\right) \partial_{R} A_{i} - \frac{1}{2} A_{i} \partial_{R} \lambda \right)\tilde{\varepsilon}^{ij}  = 0,\\
\label{CSB-EqsMotion-Boundarycond} \delta \lambda &:& - 2 \left( A_{i} \partial_{R} A_{j} \right)  \tilde{\varepsilon}^{ij} =0,\label{CSB-EqsMotion-Boundarycond}
\end{eqnarray}
if the \emph{boundary term},
\begin{equation}
\nonumber \delta S_{CSB} [A_{R}, A_{i}, \lambda] \left|_{on\,\,shell} =  \frac{\kappa}{2} \int_{\mathcal{M}} \left[ \partial_{i} \left( A_{R} \delta A_{j}- A_{j} \delta A_{R}  \right) - \frac{1}{2} \partial_{R} \left( (1 + \lambda) (A_{i} \delta A_{j} -  A_{j} \delta A_{i} )\right) \right] \tilde{\varepsilon}^{ij} \mathrm{d} R \right. ,
\end{equation}
vanishes under appropriate boundary conditions. These equations of motion describe the dynamics on each time-like surface with $\lambda$ being a \emph{label} for each member of the uniparametric family of time-like surfaces that foliate our region of space-time. And, because we have imposed the condition of the restriction to the boundary, there is no propagation of degrees of freedom from one surface $\mathcal{B}_{R}$, to the next one, $\mathcal{B}_{R + \delta R}$.

In adapted coordinates to the time-like foliation, $(t,R,\bar{x})$, this boundary term can be written as,
\begin{eqnarray}
\nonumber \delta S_{CSB} [A_{R}, A_{t}, A_{\bar{x}}, \lambda] \left|_{on\,\,shell} \right. &=&  \frac{\kappa}{2} \int_{\mathcal{M}} \left[ \partial_{t} \left( A_{R} \delta A_{\bar{x}}- A_{\bar{x}} \delta A_{R}  \right) - \partial_{\bar{x}} \left( A_{R} \delta A_{t}- A_{t} \delta A_{R}  \right) \right.  \\
&& \left.  -  \partial_{R} \left( (1 + \lambda) \left( A_{t} \delta A_{\bar{x}} -  A_{\bar{x}} \delta A_{t} \right)\right) \right] \tilde{\varepsilon}^{t\bar{x}}\, \mathrm{d} R .
\end{eqnarray}
Solving \emph{the condition of the restriction to the boundary} automatically makes the last term of the right-hand side to vanish. So the solution of this condition can be viewed as \emph{boundary condition}\footnote{Note that in the case of asymptotically flat space-times the appropriate fall-off conditions on the fields satisfy the condition of the restriction to the boundary. }. In what follows we refer to the solutions of this condition as \emph{solution to the condition of the restriction to the boundary}. And the remaining boundary term is,
\begin{equation}\label{Boundary-term-after-solving-Condition-restriction-boundary}
\delta S_{CSB}  \left|_{on\,\,shell} \right. =  \frac{\kappa}{2} \int_{\mathcal{M}} \left[ \partial_{t} \left( A_{R} \delta A_{\bar{x}}- A_{\bar{x}} \delta A_{R}  \right) - \partial_{\bar{x}} \left( A_{R} \delta A_{t}- A_{t} \delta A_{R}  \right) \right] \tilde{\varepsilon}^{t\bar{x}}\, \mathrm{d} R .
\end{equation}

After finding the \emph{solutions to the condition of the restriction to the boundary}. The action principle is still not differentiable, so we need to look for additional boundary conditions. Here is a clear distinction between the \emph{solution to the condition of the restriction to the boundary} and these additional conditions we need to impose. The first one came directly from the action principle just asking for a time-like boundary, while the later will be sensible to both the particular topology of the boundary as well as the solutions to the equations of motion, since we demand compatibility between the bulk and boundary dynamics. So we made or analysis by cases depending on the possible topologies and configurations.

\subsubsection{Topology of the time-like surface $\mathcal{ B}_{R}= S^{1} \times \mathbb{R}$}

If we identify $\bar{x} = \theta$, and remembering that $S^{1}$ does not have boundary, equation (\ref{Boundary-term-after-solving-Condition-restriction-boundary}) becomes,
\begin{equation}\label{CSB-Boundary-Term-Topology-S1xR}
\delta S_{CSB}  \left|_{on\,\,shell} \right. =  \frac{\kappa}{2} \int_{\mathcal{M}} \left[ \partial_{t} \left( A_{R} \delta A_{\bar{x}}- A_{\bar{x}} \delta A_{R}  \right)  \right] \tilde{\varepsilon}^{t\bar{x}}\, \mathrm{d} R .
\end{equation}
To make this boundary term to vanish we have the following possible additional boundary conditions:
\begin{itemize}
	\item $\mathbf{\delta A_{\bar{x}}|_{t_{1}} ^{t_{2}} = 0 = \delta A_{R}|_{t_{1}} ^{t_{2}}}$, these are the standard Dirichlet boundary conditions.
	\item \textbf{Adding a counter term to the action}, $- \displaystyle  \frac{\kappa}{2} \int_{\mathcal{M}} \left[ \partial_{t} \left( A_{R} A_{\bar{x}} \right)  \right] \tilde{\varepsilon}^{t\bar{x}}\, \mathrm{d} R $, so that under the variation we have,
	\begin{equation}
	\delta S_{CSB}  \left|_{on\,\,shell} \right. = - \kappa \int_{\mathcal{M}} \left[ \partial_{t} \left(  A_{\bar{x}} \delta A_{R}  \right)  \right] \tilde{\varepsilon}^{t\bar{x}}\, \mathrm{d} R .
	\end{equation}
	And since one of our solutions is $A_{R} = A_{R}[R]$ (case 2 in Sec. \ref{CSB-Solution-At-vAx-0}), for each time-like boundary. We can assume $\delta A_{R}|_{t_{1}} ^{t_{2}} = 0 $ without loss of generality for those solutions.
	\item $\mathbf{A_{\bar{x}} |_{t_{1},x_{1}}  = w  A_{R}|_{t_{1},x_{1}} \,\,\, with \,\,\, w = const}$. It is enough to ask that $A_{\bar{x}}$ and $A_{R}$ coincide at the initial time, $t_{1}$, and position, $x_{1}$. And by the existence and uniqueness theorem, once the initial conditions are chosen, there is a unique solution to the equations of motion compatible. This boundary/initial condition is compatible with the solution: case 1 in Sec. \ref{CSB-Solution-At-vAx-0}.
	\item  \textbf{Periodic boundary conditions:} If we assume that $A_{R} (t) = A_{R} (t + b)$ and $A_{\bar{x}} (t) = A_{\bar{x}} (t + b)$, for a region of space-time where $t \in [t,t+b]$. This can be interesting for solutions of wave propagation.
	\item \textbf{Parity conditions:} If we want the boundary term (\ref{CSB-Boundary-Term-Topology-S1xR}) to vanish under parity conditions, the term $A_{R} \delta A_{\bar{x}}- A_{\bar{x}} \delta A_{R}$, must be even in $t$, so $\partial_{t} \left( A_{R} \delta A_{\bar{x}}- A_{\bar{x}} \delta A_{R}  \right)$ is odd in $t$. Remembering that the variation does not change parity on the space-time coordinates, both $A_{R}$ and $A_{\bar{x}}$ must be even or odd in $t$ at the same time. However, if we want to have the chiral wave equation as equation of motion, it will be necessary to ask for parity of both $x$ and $t$ at the same time, such that the solutions are even or odd on the argument $(x + vt)$. This scenario, where we the introduction of parity condition is needed, has been discussed in the context of other theories: Electrodynamics \cite{Henneaux-Troessaert-2018electro}, General Relativity \cite{Ashtekar-Engle-Sloan-2008, Compere-2011, Henneaux-Troessaert-2018ADM}.
\end{itemize}

\subsubsection{Topology of the time-like surface $\mathcal{ B}_{R}=  \mathbb{R}^{2}$}

So to make the action principle well posed we need to check whether the remaining boundary term, 
\begin{equation}\label{CSB-Variation-Boundaryterm-TopologyR2}
\delta S_{CSB}  \left|_{on\,\,shell} \right. =  \frac{\kappa}{2} \int_{\mathcal{M}} \left[ \partial_{t} \left( A_{R} \delta A_{\bar{x}}- A_{\bar{x}} \delta A_{R}  \right) - \partial_{\bar{x}} \left( A_{R} \delta A_{t}- A_{t} \delta A_{R}  \right) \right] \tilde{\varepsilon}^{t\bar{x}}\, \mathrm{d} R 
\end{equation}
vanishes or if we need to impose additional boundary conditions or add a non-covariant counter-term. We have the following possibilities:
\begin{itemize}
	\item $\mathbf{\delta A_{\bar{x}}|_{t_{1}} ^{t_{2}} = 0 = \delta A_{R}|_{t_{1}} ^{t_{2}}}$, these are the standard Dirichlet boundary conditions. With this we only have the second term of equation (\ref{CSB-Variation-Boundaryterm-TopologyR2})
	\begin{itemize}
		\item \textbf{Adding a counter term to the action}, $ \displaystyle \frac{\kappa}{2} \int_{\mathcal{M}} \left[ \partial_{\bar{x}} \left( A_{R} A_{t} \right)  \right] \tilde{\varepsilon}^{t\bar{x}}\, \mathrm{d} R $, so that under the variation we have,
		\begin{equation}
		\delta S_{CSB}  \left|_{on\,\,shell} \right. =  \kappa \int_{\mathcal{M}} \left[ \partial_{\bar{x}} \left(  A_{t} \delta A_{R} \right)   \right] \tilde{\varepsilon}^{t\bar{x}}\, \mathrm{d} R .
		\end{equation}
		Which is compatible with one of our solutions $A_{R} = A_{R}[R]$ (case 2 in Sec. \ref{CSB-Solution-At-vAx-0}), for each time-like boundary. We can assume $\delta A_{R}|_{t_{1}} ^{t_{2}} = 0 $ without loss of generality for those solutions.
		
		\item $\mathbf{A_{t} |_{x_{1}} ^{x_{2}}  = w  A_{R}|_{x_{1}} ^{x_{2}} \,\,\, with \,\,\, w = const}$. Since we are using the solutions of \emph{the condition of the restriction to the boundary}, $A_{t} - v A_{\bar{x}} =0 $, $A_{R}$ is proportional to $A_{\bar{x}}$ too, and this case is compatible with the solution: case 1 in Sec. \ref{CSB-Solution-At-vAx-0}.
		\item  \textbf{Periodic boundary conditions:} We can assume that the fields are periodic in $\bar{x}$, that is, $A_{R} (\bar{x}) = A_{R} (\bar{x} + a)$ and $A_{t} (\bar{x}) = A_{t} (\bar{x} + a)$. These are the boundary conditions usually used in the Quantum Hall Effect, as we shall discuss in following sections. These conditions are widely used in Condensed Matter systems.
	\end{itemize}

	\item \textbf{Adding a counter term to the action}, $ \displaystyle \frac{\kappa}{2} \int_{\mathcal{M}} \left[ - \partial_{t} \left( A_{R} A_{\bar{x}} \right) + \partial_{\bar{x}} \left( A_{R} A_{t} \right)  \right] \tilde{\varepsilon}^{t\bar{x}}\, \mathrm{d} R$, so that under the variation we have,
	\begin{equation}
	\delta S_{CSB}  \left|_{on\,\,shell} \right. =  \kappa \int_{\mathcal{M}} \left[ - \partial_{t} \left(  A_{\bar{x}} \delta A_{R}  \right) + \partial_{\bar{x}} \left(  A_{t} \delta A_{R} \right)  \right] \tilde{\varepsilon}^{t\bar{x}}\, \mathrm{d} R .
	\end{equation}
	And since one of our solutions is $A_{R} = A_{R}[R]$ (case 2 in Sec. \ref{CSB-Solution-At-vAx-0}), for each time-like boundary. We can assume $\delta A_{R}|_{t_{1}} ^{t_{2}} = 0 $ without lost of generality for those solutions.
	\item $\mathbf{A_{\bar{x}} |_{t_{1},x_{1}}  = w  A_{R}|_{t_{1},x_{1}} \,\,\, with \,\,\, w = const}$. We are also using the solutions of \emph{the condition of the restriction to the boundary}, $A_{t} - v A_{\bar{x}} =0$, so $A_{\bar{x}}$ and $A_{t}$ are proportional too.  It is enough to ask that $A_{\bar{x}}$ and $A_{R}$ coincide at the initial time, $t_{1}$, and position, $x_{1}$. And by the existence and uniqueness theorem, once the initial conditions are chosen, there is a unique solution to the equations of motion compatible. This boundary/initial condition is compatible with the solution: case 1 in Sec. \ref{CSB-Solution-At-vAx-0}.
	\item  \textbf{Periodic boundary conditions:} This time we assume that the fields are periodic in both variables $\bar{x}$ and $t$, $A_{R} (t,\bar{x}) = A_{R} (t+b,\bar{x} + a)$ and $A_{t} (t, \bar{x}) = A_{t} (t+b,\bar{x} + a)$. This may be interesting for solutions of wave propagation.
	\item \textbf{Parity conditions:} If we want the boundary term (\ref{CSB-Boundary-Term-Topology-S1xR}) to vanish under parity conditions, the term $A_{R} \delta A_{\bar{x}}- A_{\bar{x}} \delta A_{R}$, must be even in both $t$ and $\bar{x}$, so $\partial_{t} \left( A_{R} \delta A_{\bar{x}}- A_{\bar{x}} \delta A_{R}  \right)$ is odd in $t$ and $\bar{x}$. We need to be careful that our fields are both even or odd at the same time, such that if we want to keep the chiral wave equation, we need that the functions are both even or odd in the argument $(x+vt)$. This scenario has been discussed in the context of other theories: Electrodynamics \cite{Henneaux-Troessaert-2018electro}, General Relativity \cite{Ashtekar-Engle-Sloan-2008, Compere-2011, Henneaux-Troessaert-2018ADM}.
\end{itemize}


\subsection{Dynamics at the boundary}\label{CSB-Dynamics boundary}

After discussing the well-posedness of the action principle, we analyze the solution to the equations of motion. 

We classify the solutions to the equations of motion depending on the possible solutions to the \emph{the condition of the restriction to the boundary}. Remember that this condition ensures that after the projection of the action to time-like surfaces, there are no degrees of freedom ``leaking'' from one time-like hypersurface to the next one. When we choose a particular time-like hypersurfaces (that we call the boundary) the dynamics is restricted to it. So we can say that we have a proper action with its dynamics defined at the boundary. Each of the solutions to \emph{the condition of the restriction to the boundary} leads to a particular set of solutions to the boundary that may not agree in the physics they describe.

Multiplying equation (\ref{CSB-EqsMotion-Gauss}) by $A_{j}$,
\begin{equation}
-2\left[ A_{j} \partial_{i} A_{R} - \left( 1 + \frac{\lambda}{2} \right) A_{j} \partial_{R} A_{i} - \frac{1}{2} A_{j} A_{i} \partial_{R} \lambda \right] \tilde{\varepsilon}^{ij} = 0,
\end{equation}
note that the second term vanishes due to equation (\ref{CSB-EqsMotion-Boundarycond}) and the last term vanishes by antisymmetry, so we have,
\begin{equation}\label{CSB-4thEOM}
-2  \left(A_{j} \partial_{i} A_{R} \right) \tilde{\varepsilon}^{ij} =0.
\end{equation}

In what follows we shall use adapted coordinates to the geometry of the boundary $(t,R,\bar{x})$\footnote{We are using $\bar{x}$ instead of just $x$ to emphasize that we are not working in a particular coordinate system yet. This is still a general coordinate system adapted to the boundary foliation. In the case we foliate by cylinders it will be convenient to use cylindrical coordinates or if we folliate by timelike planes we can use cartesian coordinates.}. We first solve \emph{the condition of the restriction to the boundary}, (\ref{CSB-EqsMotion-Boundarycond}). In the coordinates, $(t,R,\bar{x})$, equation (\ref{CSB-4thEOM}) can be written as,
\begin{equation}\label{CSB-4thEOM-coordinates}
- 2 \left(A_{\bar{x}} \partial_{t} A_{R} - A_{t} \partial_{\bar{x}} A_{R}  \right)  = 0.
\end{equation}

\subsubsection{Solutions to the condition of the restriction to the boundary}\label{Subsection-solution-condition-restriction-boundary}

Sometimes from physical reasons it may be clear which boundary conditions to impose, but in other cases this may not be that easy. This condition, equation (\ref{CSB-EqsMotion-Boundarycond}), ensures that we do not have degrees of freedom propagating from a timelike surface $\mathcal{B}_{R}$ to the next $\mathcal{B}_{R+\delta R}$. So, in that sense, is a \emph{boundary condition} restricting the dynamics to a hypersurface $\mathcal{B}_{R}$. 

To solve the equations of motion we first solve \emph{the condition of the restriction to the boundary},
\begin{equation*}
- 2 \left( A_{i} \partial_{R} A_{j} \right)  \tilde{\varepsilon}^{ij} =0.
\end{equation*}

The general solution of this condition, (\ref{CSB-EqsMotion-Boundarycond}), is,
\begin{equation}\label{CSB-BoundaryCondition-At-vAx}
A_{t} - v A_{\bar{x}}=0.
\end{equation}
with $v=v(t,\bar{x})$ an arbitrary function only of $t$ and $\bar{x}$.

We also have a ``trivial choice'' that make the expressión (\ref{CSB-EqsMotion-Boundarycond}) to vanish, $A_{t} = 0$, and it is widely used as a gauge condition in the bulk. Though as we shall see it is not a good boundary condition. 

And with each of these choices we may have different dynamics at the boundary, showed in the solutions to the equations of motion.

\subsubsection{Solution when $A_{t} - v A_{\bar{x}}=0$ }\label{CSB-Solution-At-vAx-0}

In the literature (see e.g. \cite{Tong-2016}), this condition has been imposed as boundary condition in order to obtain the edge modes for the Chern-Simons theory, though the authors in  \cite{Tong-2016} considered $v$ as a parameter (that is a particular case of the most general solution to eq. \ref{CSB-EqsMotion-Boundarycond}, where $v=v[t,\bar{x}]$), that will turn out to be the velocity of excitations on the boundary. In their case, they have to introduce this condition by hand, in our analysis it is part of the solution of \emph{the condition of the restriction to the boundary}.

Substituting the condition (\ref{CSB-BoundaryCondition-At-vAx}) into equation (\ref{CSB-4thEOM-coordinates}) we have,
\begin{equation}
\partial_{t} A_{R} - v \partial_{\bar{x}} A_{R}  =0,
\end{equation}
which is the chiral wave equation. On the other hand, a solution of equation (\ref{CSB-EqsMotion-F-0}) is
\begin{equation}
A_{t} = \partial_{t} \phi ,\,\,\,\, \mathrm{and} \,\,\,\, A_{\bar{x} } = \partial_{\bar{x}} \phi,
\end{equation}
where $\phi$ is valued in the Lie Algebra of $U(1)$. Considering, again, the condition (\ref{CSB-BoundaryCondition-At-vAx}), we can see that $\phi$ satisfies,
\begin{equation}
\partial_{t} \phi - v \partial_{\bar{x}} \phi=0,
\end{equation}
which is, again, the chiral wave equation.

So far, we have found the following equations, 
\begin{eqnarray}
\label{CSB-Chiral-equation-AR}  \partial_{t} A_{R} - v \partial_{\bar{x}} A_{R}  &=& 0\\
\label{CSB-Chiral-equation-phi} \partial_{t} \phi - v \partial_{\bar{x}} \phi &=& 0,
\end{eqnarray}
whose general solutions are of the form,
\begin{eqnarray}
A_{R} &=& C_{R}[R] A[x+vt]\\
\phi &=& C_{\phi}[R] \varphi [x+vt],
\end{eqnarray}
where $A[x+vt]$ and $\varphi [x+vt]$ are arbitrary functions with argument $(x+vt)$. However, waves propagating in the other direction, $A[x - vt]$ and $\varphi [x - vt]$, respectively, are not solutions.

From the first equation, (\ref{CSB-Chiral-equation-AR}), we have two possible solutions:
\begin{itemize}
	\item The general solution to the chiral wave equation, $A_{R} = C_{R}[R] A[x+vt]$.
	\item And a trivial solution\footnote{Despite being a trivial solution, we are considering it here because it is widely used in the literature (see for instance, \cite{Grumiller-Merbis-Riegler-2017}, and references therein).}, $A_{R} = A_{R}[R]$. 
\end{itemize}
So we have two sets of solutions, that we shall substitute in equations (\ref{CSB-EqsMotion-Gauss}) to understand the role of the variable $\lambda$.

\textbf{Case 1:} $A_{R} = C_{R}[R] A[x+vt]$ \textit{and} $\phi = C_{\phi}[R] \varphi [x+vt]$.

If we want the action principle to be well posed without imposing that the variation of the fields vanish at the boundary, we have several possibilities (as shown in the previous section), although not all the possible boundary conditions are compatible with all the solutions to the equations of motion. The condition that we consider is $A_{\bar{x}} |_{t_{1},x_{1}}  =  A_{R}|_{t_{1},x_{1}}$,  that implies that the solutions for $A_{R}$ and $A_{\bar{x}}$ coincide, which implies,
\begin{equation}
A_{R} = A_{\bar{x}} = C[R] A[x+vt] .
\end{equation}
Substituting into equations (\ref{CSB-EqsMotion-Gauss}) we have,
\begin{equation}
-C[R] \partial_{x} A[x+vt] + \left( 1 + \frac{\lambda}{2} \right) A[x+vt] \partial_{R} C[R] + \frac{1}{2} C[R] A[x+vt] \partial_{R} \lambda = 0.
\end{equation}
To understand the role of $\lambda$, that we introduce to enforce the condition of the restriction to the boundary, it is convenient to chose,
\begin{equation}\label{CSB-EOM-choice-exp}
A[x+vt] = e^{-(x+vt)}.
\end{equation}

We can substitute this choice for $A_{R}$ and $\phi$ into (\ref{CSB-EqsMotion-Gauss}) to see the equation satisfied by $\lambda$,
\begin{equation}
e^{-(x+vt)} \left[  C[R]+ \left( 1 + \frac{\lambda}{2} \right) \partial_{R} C[R]+ \frac{1}{2} C[R] \partial_{R} \lambda \right] =0,
\end{equation}
whose solution is,
\begin{equation}
\lambda = -2 \left(\frac{ \int_{1}^{R} C[u] \mathrm{d}u  - const}{C[R]} + 1 \right).
\end{equation}
Where, as usual, $const$ depends on the initial conditions once the particular form of $C[R]$ is chosen. Note that up to this point we have made the choice (\ref{CSB-EOM-choice-exp}), with the aim to make clear the role of $\lambda$. Which, at the end, is only a function of $R$ that can be thought of as a \emph{label} for each of the time-like surfaces of the foliation, $\mathcal{B}_{R}$.

\textbf{Case 2:} $A_{R} =  A_{R}[R]$ \textit{and} $\phi = C_{\phi}[R] \varphi [x+vt]$.

Using the condition $A_{t} - v A_{\bar{x}}=0$, the solution to (\ref{CSB-EqsMotion-F-0}), $A_{i} = \partial_{i}\phi$, and substituting them in equations (\ref{CSB-EqsMotion-Gauss}), we have,
\begin{equation}
\partial_{\bar{x}} \varphi[x+vt] \left[ \left( 1 + \frac{\lambda}{2}\right) \partial_{R} C_{\phi}[R] - \frac{1}{2} C_{\phi}[R] \partial_{R} \lambda  \right]=0,
\end{equation}
whose solution is,
\begin{equation}
\lambda [x , R] = -2 + C_{\phi}[R] f[x],
\end{equation}
where $f(x)$ is an arbitrary function, that we can set constant since it does not play any role in the dynamics of the system. From the previous equation we can see that $\lambda=\lambda[R]$ can be seen as a \emph{label} for each of the time-like surfaces of the foliation, $\mathcal{B}_{R}$.

\subsubsection{Solution when $A_{t} = 0$ }

From equations (\ref{CSB-EqsMotion-F-0}) and (\ref{CSB-4thEOM-coordinates}), and the condition $A_{t} = 0$, we have that 
\begin{equation}
\partial_{t} A_{\bar{x}} = 0 \,\,\,\, \text{and} \,\,\,\,  \partial_{t} A_{R} = 0.
\end{equation}
Whose solutions are,
\begin{equation}
A_{\bar{x}} = A_{\bar{x}} [\bar{x} , R]	 \,\,\,\, \text{and} \,\,\,\, A_{R} = A_{R} [\bar{x} , R],
\end{equation}
respectively. From the two equations (\ref{CSB-EqsMotion-Gauss}), the second equation is the only non-vanishing one,
\begin{equation}
\partial_{\bar{x}} A_{R} - \left(1 + \frac{\lambda}{2}\right) \partial_{R} A_{\bar{x}} - \frac{1}{2} A_{\bar{x}} \partial_{R} \lambda = 0.
\end{equation}
Since the condition, $A_{t} = 0$, satisfy trivially \emph{the condition of the restriction to the boundary}, it seems that we do not have enough equations for our fields. Even if we introduce some of the boundary conditions needed for the action principle to be well posed, we cannot uniquely determine the degree of freedom nor give $\lambda$ a meaning. So this solution to the condition of the restriction to the  boundary is not well suited for our purposes. 

Note that even though $A_{t}=0$ is a good gauge fixing in the bulk, it is not useful as boundary condition. The corresponding equations of motion have solutions that do not depend on $t$. 

\subsection{The reduced action at the boundary}\label{Section-reduced-actions-in-the-boundary}

As we shall show later, there is a correspondence between the gauge fixing conditions in the bulk with the solutions to \emph{the condition of the restriction to the boundary}, which can be seen as boundary condition. So loosely speaking we can say that there is a correspondence between the gauge fixing at the bulk and boundary conditions. We clarify this point in Section (\ref{Section-HamiltonianAnalysis-Bulk-vs-Boundary}) after having all the ingredients to make the statements more precise.

The action (\ref{Chern-Simons-Boundary}) can be written in adapted coordinates $(t,R,\bar{x})$ as,
\begin{eqnarray}\label{CS-CompleteAction-Boundary-tRx}
\nonumber	S_{CSB} [A_{t}, A_{R}, A_{\bar{x}},\lambda] &=& \frac{\kappa}{4} \int_{ \mathcal{ M}} \left[ 2 A_{R} \left( \partial_{t} A_{\bar{x} } -\partial_{\bar{x}} A_{t} \right) + 2 \left( A_{t} \partial_{\bar{x}} A_{R} - A_{\bar{x}} \partial_{t} A_{R} \right) \right.\\
&& \left. -2 \left( 1 + \lambda \right)\left( A_{t} \partial_{R} A_{\bar{x}} - A_{\bar{x}} \partial_{R} A_{t} \right) \right] \tilde{\varepsilon}^{t\bar{x} } \, \mathrm{d} R.
\end{eqnarray}
If the interest is just to characterize the degrees of freedom and their dynamics at the boundary, it may be easier to first solve \emph{the condition of the restriction to the boundary} and reduce the theory\footnote{As we show below at Sec. (\ref{Section-HamiltonianAnalysis-Bulk-vs-Boundary}), the theory at the boundary only has second class constraints. It is equivalent to solve the second class constraints and reduce the system and then to study its properties, chap. 12 of \cite{Henneaux-Teitelboim-1994}.}.

Solving \emph{the condition of the restriction to the boundary} we found in section \ref{Subsection-solution-condition-restriction-boundary} that a particular solution of this constraint is, $A_{t} - v A_{\bar{x}} =0$ with $v=const$. Substituing this into the previous action (\ref{CS-CompleteAction-Boundary-tRx}),
\begin{equation}\label{Reduced-action-At-vAx-v-const}
S_{R_1} = \frac{\kappa}{2} \int_{ \mathcal{ M}} \left[ A_{R} \left( \partial_{t} A_{\bar{x}} - v \partial_{\bar{x}} A_{\bar{x}} \right) - A_{\bar{x}} \left( \partial_{t} A_{R}  - v \partial_{\bar{x}} A_{R} \right) \right]\tilde{\varepsilon}^{t\bar{x}}
\end{equation}
Taking the variation of this action
\begin{eqnarray}
\nonumber \delta	S_{R_1} &=&  \frac{\kappa}{2} \int_{ \mathcal{ M}} \left[  \partial_{t} \left( A_{R} \delta A_{\bar{x}} - A_{\bar{x}} \delta A_{R} \right) + v \partial_{\bar{x}} \left( A_{\bar{x}} \delta A_{R} - A_{R} \delta A_{\bar{x}} \right) \right. \\
&& \left. + 2\delta A_{R} \left( \partial_{t} A_{\bar{x}} - v \partial_{\bar{x}}  A_{\bar{x}} \right) - 2 \delta A_{\bar{x}} \left( \partial_{t} A_{R} - v \partial_{\bar{x}} A_{R} \right)  \right]\tilde{\varepsilon}^{t\bar{x}},
\end{eqnarray}
using the boundary conditions discussed above, the equations of motion are,
\begin{eqnarray}
\nonumber \partial_{t} A_{\bar{x}} - v \partial_{\bar{x}}  A_{\bar{x}} &=& 0\\
\nonumber \partial_{t} A_{R} - v \partial_{\bar{x}} A_{R} &=& 0.
\end{eqnarray}
Which are the chiral wave equations and whose general solutions are,
\begin{eqnarray}
A_{\bar{x}} &=& C_{\bar{x}} \varphi_{\bar{x}} [x+vt]\\
A_{R} &=& C_{R} \varphi_{R} [x+vt],
\end{eqnarray}
where $C_{\bar{x}}$ and $C_{R}$ are constants (that may depend only on the fixed chosen value of $R$), and $\varphi_{\bar{x}} [x+vt]$ and $\varphi_{R} [x+vt]$ are arbitrary functions with argument $[x+vt]$.
As shown in the previous section, one possibility to make the complete action principle (\ref{Chern-Simons-Boundary}) well posed is the condition,
\begin{equation}
A_{\bar{x}} |_{t_{1},x_{1}}  = w  A_{R}|_{t_{1},x_{1}}.
\end{equation} 
This condition also ensures that the boundary term in the variation of this reduced action principle vanishes, so it is differentiable.
If the solutions to the equations of motion coincide at initial positions and times, by the existence and uniqueness theorem, we can say that the solutions coincide,
\begin{equation}
A_{\bar{x}} =  A_{R} = C \varphi [x+vt]
\end{equation}
The reduced action principle (\ref{Reduced-action-At-vAx-v-const}) describes a theory with one degree of freedom, characterized by $\varphi [x+vt]$, such that obeys the chiral wave equation.

\subsection{Well-posedness of the bulk action}

For the three-dimensional Chern-Simons theory in the bulk, eq. (\ref{CSAction-components}), the variation of the action is (see eq. \ref{CSBulk-Boundaryterm}),
\begin{equation}
\frac{\kappa}{2} \int_{ \mathcal{M}} \left[ \partial_{i} \left( A_{R} \delta A_{j}- A_{j} \delta A_{R}  \right)  - \frac{1}{2} \partial_{R} \left( A_{i} \delta A_{j} - A_{j} \delta A_{i} \right)  \right] \tilde{\varepsilon}^{ij} \mathrm{d} R.
\end{equation}
In adapted coordinates to the time-like foliation, $(t,R,\bar{x})$, this boundary term can be written as,
\begin{eqnarray}
\nonumber \delta S_{CSB} [A_{R}, A_{t}, A_{\bar{x}}, \lambda] \left|_{on\,\,shell} \right. &=&  \frac{\kappa}{2} \int_{\mathcal{M}} \left[ \partial_{t} \left( A_{R} \delta A_{\bar{x}}- A_{\bar{x}} \delta A_{R}  \right) - \partial_{\bar{x}} \left( A_{R} \delta A_{t}- A_{t} \delta A_{R}  \right) \right.  \\
&& \left.  -  \partial_{R}  \left( A_{t} \delta A_{\bar{x}} -  A_{\bar{x}} \delta A_{t} \right) \right] \tilde{\varepsilon}^{t\bar{x}}\, \mathrm{d} R .
\end{eqnarray}
Note that as well as for the boundary action, the solutions to \emph{the condition of the restriction to the boundary} make the last term to vanish. And the additional boundary conditions discussed in Sec. \ref{Section-reduced-actions-in-the-boundary}, also make the action principle well posed. And the choice, of course, will depend on the allowed solutions to the equations of motion we want to study. In this way there is a complete agreement between the dynamics in the bulk and the boundary, as expected.

\subsection{Hamiltonian analysis bulk vs  boundary}\label{Section-HamiltonianAnalysis-Bulk-vs-Boundary}

In this part, we shall highlight the main results from the Hamiltonian analysis both in the bulk and the boundary to be able to compare them. We left the details of the calculations to the appendices.

In what follows we use the standard Hamiltonian analysis \cite{Henneaux-Teitelboim-1994}. To start we need to define an action principle. From the bulk action we can read \emph{the condition of the restriction to the boundary} so that there are no degrees of freedom propagating from one time-like hypersurface to the next one, from which we construct the corresponding theory at the boundary. Both actions are defined on the same region of the space-time, but the boundary one contains the information of the dynamics on the family of time-like hypersurfaces, each being independent from one another (see table \ref{table-Action principles}).
\begin{table}[ht]
	\begin{center}
		\caption{Action principles}
		\label{table-Action principles}
		\begin{tabular}{l|l}
			\toprule 
			\textbf{Bulk} & $S_{CS} [\mathbf{A}] =  \displaystyle \frac{\kappa}{4} \int_{\mathcal{M}} \left[ (A_{\rho} R^{\rho}) F_{\mu \nu} + 2 A_{\mu} \partial_{\nu} (A_{\rho} R^{\rho}) - 2 A_{\mu} \pounds_{\vec{R}} A_{\nu} \right] \tilde{\varepsilon}^{ \mu \nu} \mathrm{d} R$ \\
			\midrule
			\textbf{Boundary} & $S_{CSB} [\mathbf{A}, \lambda]= \displaystyle \frac{\kappa}{4} \int_{\mathcal{M}} \left[ (A_{\rho} R^{\rho}) F_{\mu \nu} + 2 A_{\mu} \partial_{\nu} (A_{\rho} R^{\rho}) - 2 (1 + \lambda) A_{\mu} \pounds_{\vec{R}} A_{\nu}   \right] \tilde{\varepsilon}^{ \mu \nu} \mathrm{d} R$ \\
			\bottomrule 
		\end{tabular}
	\end{center}
\end{table}

Note that in the boundary action we have introduced an additional auxiliary dynamical variable, $\lambda$, that enforces \emph{the condition of the restriction to the boundary}. Now the number of our phase space variables, instead of $6$ are $8$. See table \ref{table:Phase space variables}, for a comparison.
\begin{table}[ht]
	\begin{center}
		\caption{Phase space variables}
		\label{table:Phase space variables}
		\begin{tabular}{l|l}
			\toprule 
			\textbf{Bulk} & \textbf{Boundary} \\
			\midrule 
			$A_{t}$, $\Pi_{t}$ & $A_{t}$, $\Pi_{t}$ \\
			$A_{R}$, $\Pi_{R}$ & $A_{R}$, $\Pi_{R}$ \\
			$A_{\bar{x}}$, $\Pi_{\bar{x}}$ & $A_{\bar{x}}$, $\Pi_{\bar{x}}$\\
			& $\lambda$, $\Pi_{\lambda}$\\
			\bottomrule 
		\end{tabular}
	\end{center}
\end{table}

From the definition of the canonical momenta we can find the primary constraints, as we can see we have an extra primary constraint associated with the new dynamical variable $\lambda$, see table (\ref{Table-Primary constraints}).
\begin{table}[ht]
	\begin{center}
		\caption{Primary constraints}
		\label{Table-Primary constraints}
		\begin{tabular}{l|l}
			\toprule 
			\textbf{Bulk} & \textbf{Boundary} \\
			\midrule 
			$\phi_{t} : = \Pi_{t} \approx 0$ & $\phi_{t} := \Pi_{t} \approx 0$\\
			$\phi_{R} :=	\Pi_{R}  + \alpha A_{\bar{x}} \approx 0$ & $\phi_{R} :=	\Pi_{R}  + \alpha A_{\bar{x}} \approx 0$ \\
			$\phi_{\bar{x}} :=	\Pi_{\bar{x}} - \alpha A_{R} \approx 0$ & $\phi_{\bar{x}} :=	\Pi_{\bar{x}} - \alpha A_{R} \approx 0$\\
			& $\phi_{\lambda} := \Pi_{\lambda} \approx 0$\\
			\bottomrule 
		\end{tabular}
	\end{center}
\end{table}

We need to check the consistency condition, we ask for the constraints to not change in time. We found the one extra secondary constraint, $\psi_{\lambda} $, that is \emph{the condition of the restriction to the boundary}. See table \ref{Table-Secondary constraints}.
\begin{table}[ht!]
	\begin{center}
		\caption{Secondary constraints}
		\label{Table-Secondary constraints}
		\begin{tabular}{l|l}
			\toprule 
			\textbf{Bulk} & \textbf{Boundary} \\
			\midrule 
			$\psi := - \partial_{\bar{x}} A_{R}  +  \partial_{R} A_{\bar{x}} \approx 0$ & $\psi_{t} :=  2 \alpha \partial_{\bar{x}} A_{R}  - 2\alpha (1 + \lambda)  \partial_{R} A_{\bar{x}} \approx 0$\\
			$\lambda_{\bar{x}} = - \partial_{\bar{x}} A_{t}$ & $\psi_{\lambda} :=  \alpha \left( A_{t} \partial_{R} A_{\bar{x}} - A_{\bar{x}} \partial_{R} A_{t} \right) \approx 0$ \\
			$\lambda_{R} = -  \partial_{R} A_{t}$ & $\lambda_{\bar{x}} =  \partial_{\bar{x}} A_{t}$\\
			& $\lambda_{R} =   (1 + \lambda)  \partial_{R} A_{t}$\\
			\bottomrule 
		\end{tabular}
	\end{center}
\end{table}

For both the bulk and boundary theories there are no tertiary constraints. After finding all the constraints we need to classify them into first and second class. And here is where it lies the main difference between the two theories: The theory in the bulk has both first and second class constraints whereas the boundary one only has second class constrains.
\begin{table}[ht!]
	\begin{center}
		\caption{First and second class constraints}
		\label{tab:table1}
		\begin{tabular}{l|l}
			\toprule 
			\textbf{Bulk} & \textbf{Boundary} \\
			\midrule
			\hline
			\multicolumn{1}{l}{First class constraints}\\
			\hline
			$\gamma_{1} := \Pi_{t} \approx 0$ & Does not have\\
			$\gamma_{2} := \partial_{R} \Pi_{R} + \partial_{\bar{x}} \Pi_{\bar{x}} + 2\alpha \left( \partial_{R} A_{\bar{x}} + \partial_{\bar{x}} A_{R} \right) \approx 0$ &  \\
			\midrule
			\hline
			\multicolumn{1}{l}{Second class constraints}\\
			\hline
			$\chi_{1} := \Pi_{R}  + \alpha A_{j} \approx 0$ & $\chi_{1} := \phi_{t} = \Pi_{t} \approx 0$\\
			$\chi_{2} := \Pi_{\bar{x}} - \alpha A_{R} \approx 0$ & $\chi_{2} := \phi_{R} =	\Pi_{R}  + \alpha A_{\bar{x}} \approx 0$ \\
			& $\chi_{3} := \phi_{\bar{x}} =	\Pi_{\bar{x}} - \alpha A_{R} \approx 0$\\
			& $\chi_{4} := \phi_{\lambda} = \Pi_{\lambda} \approx 0$\\
			& $\chi_{5} := \psi_{t} =  2 \alpha \partial_{\bar{x}} A_{R}  - 2\alpha (1 + \lambda)  \partial_{R} A_{\bar{x}} \approx 0$\\
			& $\chi_{6} := \psi_{\lambda} = \alpha \left( A_{t} \partial_{R} A_{\bar{x}} - A_{\bar{x}} \partial_{R} A_{t} \right) \approx 0$\\
			\bottomrule 
		\end{tabular}
	\end{center}
\end{table}

While the bulk theory is a gauge theory (in fact, is a topological one that is `pure gauge'), the boundary theory does not have first class constraints (generators of gauge transformations), therefore is not a gauge theory. It only has second class constraints as shown in table \ref{Table-Gauge symmetries}.
\begin{table}[ht]
	\begin{center}
		\caption{Gauge symmetries}
		\label{Table-Gauge symmetries}
		\begin{tabular}{l|l}
			\toprule 
			\textbf{Bulk} & \textbf{Boundary} \\
			\midrule 
			$A'_{\mu} = A_{\mu} + \partial_{\mu} \varepsilon$ & Does not have \\
			\bottomrule 
		\end{tabular}
	\end{center}
\end{table}

The introduction of \emph{the condition of the restriction to the boundary}, that we already showed can be seen as boundary condition, can be interpreted as a partial gauge fixing in the bulk. So the emergence of the degree of freedom at the boundary can be seen a coming from breaking the gauge symmetry of the system. In this sense we can talk about a \emph{correspondence} between a  boundary condition and a partial gauge fixing at the bulk. However, we did not take that path in this work (constructing the boundary theory by a partial gauge fixing as discussed in other works \cite{Elitzur-etal-1989,Geiller-2017,Tong-2016,Troessaert-2013}), we start from the bulk action and only taking geometrical considerations we constructed the theory at the boundary, and we are able to show that the results are in agreement with those in the literature. The discussion on how to fix the gauge is not a trivial one, so we present an alternative approach, we read the condition of the restriction to the boundary directly from the action, and at the end we show its relation with a partial gauge fixing.
Another important difference is the number the degrees of freedom, while the bulk theory is topological (zero local degrees of freedom), the boundary has one degree of freedom, that we can associate with a chiral boson as has been found through different approaches (see Section \ref{Section-QuatumHallEffect} for the discussion).
\begin{table}[ht]
	\begin{center}
		\caption{Degrees of freedom $= \frac{1}{2} \left[(phase\,\, space \,\, variables) - 2 ( 1st-class) - (2nd-class)\right]$}
		\label{tab:table1}
		\begin{tabular}{l|l}
			\toprule 
			\textbf{Bulk} & \textbf{Boundary} \\
			\midrule 
			$D.O.F.= \frac{1}{2} \left[(6) - 2 (4) - (2)\right]= \mathbf{0}$ & $D.O.F.= \frac{1}{2} \left[(8) - 2 (0) - (6)\right] = \mathbf{1}$\\
			\bottomrule 
		\end{tabular}
	\end{center}
\end{table}

\subsection{Edge modes: Quantum Hall Effect}\label{Section-QuatumHallEffect}

In this section, we shall show how this method can be used to recover known results without the need for further assumptions. It is important to emphasize that this is an improvement from current methods, which need to introduce several conditions by hand, where for this example of the abelian Chern-Simons, it is simple enough to see which conditions can be added. But for more complicated theories, like Gravity whose boundary lies at infinity, may not be clear what to impose without losing crucial information. So to have an alternative general procedure is a useful addition to the current methods.

Here, our primary references will be two excellent lecture notes \cite{Dunne-1999,Tong-2016} to be able to compare our results with the existing ones and show their agreement.

\subsubsection{Effective Action}

It is known that for a quantum Hall fluid confined to a finite region there will be gapless modes that live on the edge. In integer quantum Hall states they are chiral, that is they propagate only in one direction. This edge modes, besides offering the best characterization for the quantum Hall states, they provide a link between the Chern-Simons approach and the microscopic wave functions  \cite{Dunne-1999,Tong-2016}.

The low-energy effective action for the Laughlin state is 
\begin{equation}\label{CS-Quantum Hall}
S_{CS} [A] = \frac{m}{4 \pi} \int_{\mathcal{M}} A_{\mu} \partial_{\nu} A_{\rho} \tilde{\epsilon}^{\mu \nu \rho},
\end{equation}
where we are working in units in which $e=\hbar=1$ and $\kappa = \frac{m}{2 \pi}$. To find the edge modes we need to analize the theory on the boundary.

The first step is to project the action (\ref{CS-Quantum Hall}) to the boundary. For that, we are considering that $\mathcal{ M}$ is a manifold with boundary $\mathcal{ B}$ that, in this simple case, can be thought of a straight plane at $y=const.=C$. The Hall state lies at $y<C$ while at $y>C$ is only the vacuum (see Fig. \ref{BoundaryHallState1})\footnote{Note that we do not even need to chose $y=0$, the approach here is the same independently of the value of $y$ chosen.}. For the geometry of the problem it is convenient to use cartesian coordinates and to analyze the problem we need a family of planes $\mathcal{ B}_{y}$, at $y=C$, with $C \in \mathbb{R}$, and at the end of the analysis we can take into account which particular surface we choose as our boundary.

\begin{figure}[h]
	\begin{center}
		\includegraphics[width=8cm]{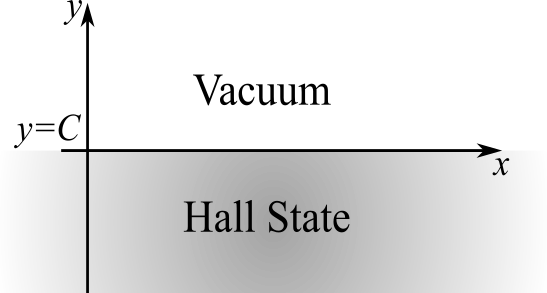}
		\caption{The boundary $\mathcal{B}$ is the straight plane at $y=C$, the time direction is coming out the page.}\label{BoundaryHallState1}
	\end{center}
\end{figure}

We can use equation (\ref{BoundaryAction-munu}), 
\begin{equation}\label{CS-projection-famil-QuantumHall}
S_{CS} [A] =  \frac{m}{8\pi} \int_{\mathcal{M}} \left[ (A_{\rho} R^{\rho}) F_{\mu \nu} + 2 A_{\mu} \partial_{\nu} (A_{\rho} R^{\rho}) - 2 (1 + \lambda) A_{\mu} \pounds_{\vec{R}} A_{\nu}   \right] \tilde{\varepsilon}^{ \mu \nu} \mathrm{d} y,
\end{equation}
considering that, in this example, our boundary is a family of hypersurfaces defined by the normal vector field $n^{\rho} = (0,0,1)$ so $A_{\rho} R^{\rho} = A_{2} =: A_{y}$, and using adapted coordinates to foliation, $\partial_{y} := \pounds_{\vec{R}}$, we obtain the theory at the boundary,
\begin{equation}\label{Chern-Simons-Boundary-QuantumHall}
S_{CSB} [A_{y}, A_{i}] =  \frac{m}{8 \pi} \int_{\mathcal{M}} \left[ A_{y}  F_{ij} + 2 A_{i} \partial_{j} A_{y} - 2 (1 + \lambda) A_{i} \partial_{y} A_{j} \right] \epsilon^{ij}  \mathrm{d} y \mathrm{d}x \mathrm{d}t, 
\end{equation}
where $\tilde{\varepsilon}^{ \mu \nu} = \epsilon^{ij} \mathrm{d}x \mathrm{d}t$ in Cartesian coordinates, $\epsilon^{ij}$ being the Levi-Civita symbol and $i,j=0,1=t,x.$

We emphasise that to obtain this theory on the boundary, we just take into account the geometry of the problem. The projection is a purely geometric one, without further fixings or assumptions. Also, we have not integrated by parts. That is one of the beauties of this method.

\subsubsection{The reduced system and the chiral boson}

We already make the general calculation for any time-like boundary, so there is no need to repeat the calculations here, for details see Appendix \ref{Hamiltonian-analysis-complete-theory}. We shall consider the reduced extended action (\ref{CSB-Reduced-extended-action})
\begin{equation}
S_{E-Reduced}[A_{x}, \Pi_{x}]= \int_{\mathcal{M}} \left[  \Pi_{x} \left( \pounds_{\vec{t}} A_{x} -  \partial_{x} \left( v(t,x) A_{x} \right)  \right) - A_{x} \left( \pounds_{\vec{t}} \Pi_{x} - v(t,x) \partial_{x} \Pi_{x} \right)    \right] \mathrm{d}x \mathrm{d}y \mathrm{d}t.
\end{equation}
and the corresponding equations of motion (\ref{CSB-Reduced-equations-motion-from-SERreduced}),
\begin{eqnarray}
\partial_{t} A_{\bar{x}} -  \partial_{\bar{x}} \left[ v(t,\bar{x}) A_{\bar{x}} \right] &=& 0\\
\partial_{t} \Pi_{\bar{x}} - v(t,\bar{x}) \partial_{\bar{x}} \Pi_{\bar{x}} &=& 0,
\end{eqnarray}
or equivalently,
\begin{eqnarray}
\partial_{t} A_{\bar{x}} -  \partial_{\bar{x}} \left[ v(t,\bar{x}) A_{\bar{x}} \right] &=& 0\\
\label{QHE-Chiralequation-AR}\partial_{t} A_{R} - v(t,\bar{x}) \partial_{\bar{x}} A_{R} &=& 0.
\end{eqnarray}

From these equations, we know from differential equations that given a set of initial/boundary conditions the solutions to the equations of motion well be unique. From Sec. (\ref{Subsection-Well-posedness-action-Boundary-conditions}), we have several possible boundary conditions that make the action principle well-posed:
\begin{itemize}
	\item {$\mathbf{A_{R} = A_{R}[R]}$} (case 2 in Sec. \ref{CSB-Solution-At-vAx-0}): In this case, equation (\ref{QHE-Chiralequation-AR}), vanish and there is only one chiral equation for the one degree of freedom of the theory at the boundary: $A_{\bar{x}}$.
	\item  $\mathbf{A_{\bar{x}} |_{t_{1},x_{1}}  = w  A_{R}|_{t_{1},x_{1}} \,\,\, with \,\,\, w = const}$: With this boundary condition, at initial $x$ and $t$, $A_{\bar{x}}$ is proportional to $A_{R}$, so the equations will coincide if $v=const.$. And for the uniqueness and existence theorem, they must coincide. And again, we have only one equation, the chiral wave equation, that characterizes the one degree of freedom.
	\item  \textbf{Periodic boundary conditions:} These are the conditions considered in \cite{Geiller-2017,Tong-2016}. Together with the particular case when $v =  const$, the one considered in the literature (see e.g. \cite{Geiller-2017,Tong-2016}), the resulting equations of motion are,
\begin{eqnarray}
\pounds_{\vec{t}} A_{x} -   v(t,x) \partial_{x} A_{x}  &=& 0\\
\pounds_{\vec{t}} \Pi_{x} - v(t,x) \partial_{x} \Pi_{x} &=& 0,
\end{eqnarray}
which are chiral wave equations. Moreover if we identify, as done in \cite{Tong-2016}, we can chose $A_{x}  =:  \phi$ and $\Pi_{x} = \frac{1}{2\pi} \partial_{x} \phi =: \rho$, the reduced extended action can be written as, 
\begin{equation}
S_{E-Reduced}[\phi]= \int_{\mathcal{M}} \left[  \frac{1}{\pi} \frac{\partial \phi}{\partial x} \left( \partial_{t} \phi - v \partial_{x} \phi  \right)  - \partial_{x} \left( \frac{\phi}{2\pi} \left( \partial_{t} \phi - v \partial_{x} \phi  \right) \right)   \right] \mathrm{d}x \mathrm{d}y \mathrm{d}t,
\end{equation}
which is equivalent to the Floreanini-Jackiw action, whose equation of motion is the chiral wave equation and the degree of freedom corresponds to the chiral boson
\begin{equation}\label{QHE-chiral-wave-equation}
\partial_{t} \phi - v \partial_{x} \phi  =0.
\end{equation}
And as it is their case, $\rho$ has the interpretation of the charge density along the boundary \cite{Tong-2016}. But note that to recover the chiral wave equation we do not need to make this choice, $A_{x}  =:  \phi$ and $\Pi_{x} = \frac{1}{2\pi} \partial_{x} \phi =: \rho$, by hand. Directly from using the allowed boundary conditions the chiral wave equation, equation (\ref{QHE-chiral-wave-equation}), emerges as one possibility to characterize the dynamics at the boundary. 
\end{itemize}

In this work we did not attempted to make a full derivation from first principles of the Quantum Hall Effect. We wanted to present a general method to construct the corresponding theory at the boundary for a field theory, and in particular we exemplify it with one of the simplest yet interesting examples, the three-dimensional abelian Chern-Simons theory. That can also be seen as the low-energy effective action for the Laughlin state of the Quantum Hall Effect, and we only show the agreement of our results with the one already in the literature for this case.

\section{Conclusions and discussion}

Given a field theory defined on a region of space-time with boundaries, given by an action principle, we provide a general method to construct the corresponding theory at the boundary with a minimal set of assumptions. The algorithm of the method goes as follows:
\begin{enumerate}
	\item Choose the action principle.
	\item Choose the type of boundary (time-like, space-like, null).
	\item Project the theory to a family of boundaries.
	\item Construct the theory at the boundary, given by an action principle.
\end{enumerate}
The idea behind the construction of the theory at the boundary is a simple yet physical and geometric: we want to restrict the degrees of freedom to the boundary, that is, we do not want to allow degrees of freedom propagating outside it. From the projected action, we can read which condition to impose as a constraint to ensure this, and we call it \emph{the condition of the restriction to the boundary}. Note that the projected bulk action tell us which condition to choose that is compatible with the action principle; otherwise if we impose any other constraint by hand we will end up with an infinite number of constraints \cite{Romero-Vergara-2002,Sheikh-Jabbari-Shirzad-2001}. As a byproduct, the solutions of the \emph{the condition of the restriction to the boundary} can be seen as boundary conditions. However, they are not enough for making the action principle well-posed, we need additional boundary conditions such as periodic or parity conditions or the addition of a counter-term, as it was pointed out in \cite{Henneaux-Troessaert-2018ADM,Henneaux-Troessaert-2018electro,Compere-2011}. We study the possible scenarios and implications.

Even though there are a lot of excellent works studying aspects of gauge theories defined on space-times with boundaries \cite{Barbero-Diaz-Margalef-Bentabol-Villasenor-2019,Brown-Henneaux-1986,Buffenoir-Henneaux-Noui-Roche-2004,Compere-2011,Corichi-Reyes-2015,Corichi-Rubalcava-Vukasinac-Review,Corichi-Vukasinac-2019,Corichi-Vukasinac-2020,Gallardo-Montesinos-2011,Geiller-2017,Grumiller-Merbis-Riegler-2017,Henneaux-Troessaert-2018ADM,Henneaux-Troessaert-2018electro,Romero-Vergara-2002,Sheikh-Jabbari-Shirzad-2001,Troessaert-2013}, as far as we know, there is no systematic approach on how to construct a theory at the boundary. From which we can extract all the relevant information at the boundary, as the gauge symmetries and especially the degrees of freedom (to count them and characterise them.) This manuscript is our contribution to filling this gap in the literature.

We apply this method to the widely studied three-dimensional abelian Chern-Simons theory. We found that the theory at the boundary is no longer a gauge theory and that has one degree of freedom characterised by the chiral boson; in contrast with the bulk theory that is topological gauge theory.  We recover these results only with the initial and minimal assumptions without any gauge fixing nor imposing additional conditions. 

\subsection*{Outlook}

The goal of this program is to construct the theory at the boundary of any field theory, to find its gauge symmetries, degrees of freedom and characterise them. We want to understand what is the mechanism of the emergence of degrees of freedom at the boundary in some topological theories in a general way. Also, we want to find a relation on how the bulk and boundary degrees of freedom are related. Furthermore, one of the open questions in the study of the field theories in space-times with boundaries is if the addition of boundary terms to the bulk action change the corresponding theory at the boundary, its symmetries and degrees of freedom. Some of these questions are currently under study, and we expect to report our findings in future works. These ideas can also be extended to a geometrical description as presented in \cite{Barbero-Diaz-Margalef-Bentabol-Villasenor-2019}.

\appendix

\section{Geometrical preliminaries}\label{Section-Geometrical-Preliminaries}

To study the Chern-Simons theory in 3-dimensions, there is no need for a metric to formulate the theory (since we are dealing with a topological theory in 3-dimensions). Also, since the edge modes of the Quantum Hall Effect, studied below, in Sec. \ref{Section-QuatumHallEffect}, are characterized using standard Cartesian coordinates (to make contact with the results in the literature), we shall motivate the geometrical tools needed in a more elementary way, the hope is that the main ideas can be follow for non-experts in differential geometry, though we shall give the formal definitions in Section \ref{Section-geometry-definitions}. 

In $\mathbb{R}^{2}$, given two linearly independent vectors $\mathbf{n}$ and $\mathbf{v}$, any other vector $\mathbf{T}$  can be written as: $\mathbf{T} = a \mathbf{n} + b \mathbf{v}$, with $a,b \in \mathbb{R}$. Also, $\mathbf{T}$ can be \emph{projected} into its components along $\mathbf{n}$ and $\mathbf{v}$. In particular, in Cartesian coordinates any vector $\mathbf{T}$ can be written as: $\mathbf{T} = T_{x} \mathbf{\hat{i}}  + T_{y} \mathbf{\hat{j}}$, where $T_{x}$ and $T_{y}$ are the $x$ and $y$ components of $\mathbf{T}$ respectively, and $\mathbf{\hat{i}}$ and $\mathbf{\hat{j}}$ are the unitary vectors determining the direction of the $x$ and $y$ axis. As it is well known, $T_{x} = \mathbf{T} \cdot \mathbf{\hat{i}} = |\mathbf{T}| \cos \theta$ and $T_{y} = \mathbf{T} \cdot \mathbf{\hat{j}} = |\mathbf{T}| \sin \theta$, where $\theta$ is the angle between the $x$-axis and the vector $\mathbf{T}$. When we want to decompose any vector in two linearly independent vectors not necessarily oriented in the direction of the coordinate axis, we follow the same idea but now the unitary vectors, $\mathbf{\hat{v}} = v^{\mu} \hat{\mathbf{e}}_{\mu}$ and $\mathbf{\hat{n}} = n^{\mu} \hat{\mathbf{e}}_{\mu}$, have components along the unit vectors $\hat{\mathbf{e}}_{\mu}$ for an arbitrary coordinate system. We chose the unitary vectors along what we choose as the \emph{tangential direction}, $\mathbf{\hat{v}}$, and the \emph{normal direction}, $\mathbf{\hat{n}}$, such that $\mathbf{\hat{v}} \cdot \mathbf{\hat{n}} = 0$.
Then we can write, $\mathbf{T} = (\mathbf{T} \cdot \mathbf{\hat{v}}) \mathbf{\hat{v}} + (\mathbf{T} \cdot \mathbf{\hat{n}})\mathbf{\hat{n}}$ in components in terms of the general coordinate unit vectors $\hat{\mathbf{e}}$ as,
\begin{equation}	
\mathbf{T} = T^{\mu} \hat{\mathbf{e}}_{\mu} = \left[ (\mathbf{T} \cdot \mathbf{\hat{v}}) v^{\mu} + (\mathbf{T} \cdot \mathbf{\hat{n}}) n^{\mu} \right] \hat{\mathbf{e}}_{\mu} = \left[  (T^{\nu} v_{\nu}) v^{\mu} +(T^{\nu} n_{\nu}) n^{\mu} \right] \hat{\mathbf{e}}_{\mu}.
\end{equation}
The components of $\mathbf{T}$ along the tangential and normal directions given by $\mathbf{\hat{v}}$ and $\mathbf{\hat{n}}$, respectively, expressed in \emph{a general coordinate system} whose unit vectors are $\hat{\mathbf{e}}_{\mu}$ satisfy,
\begin{equation}
T^{\mu} =  (T^{\nu} v_{\nu}) v^{\mu} +(T^{\nu} n_{\nu}) n^{\mu}.
\end{equation}
We can write the component along the tangential direction in terms of the components of the vector and its component along the normal direction as,
\begin{equation}
\underbrace{ v_{\nu} v^{\mu} T^{\nu}}_{Tangential \,\,part} = T^{\mu} - \underbrace{T^{\nu} n_{\nu} n^{\mu}}_{Normal \,\,part} = \delta^{\mu} _{\nu} T^{\nu} - n_{\nu} n^{\mu} T^{\nu},
\end{equation}
which can also be written as,
\begin{equation}
P^{\mu} _{\nu} T^{\nu} := (v_{\nu} v^{\mu}) T^{\nu}  = (\delta^{\mu} _{\nu}  - n_{\nu} n^{\mu}) T^{\nu}.
\end{equation}
In the previous equation we have defined the 2-dimensional \emph{projector} in $\mathbb{R}^{2}$,  $P^{\mu} _{\nu} := \delta^{\mu} _{\nu}  - n_{\nu} n^{\mu}$, to the tangential direction given by $\mathbf{\hat{v}}$, in terms of the components of the vector and its normal part in an arbitrary coordinate system.

Note that while the choice of  $\mathbf{\hat{v}}$ and $\mathbf{\hat{n}}$ is inspired by the geometry of the system, we can still work with arbitrary coordinates. It is only when we chose an adapted coordinate system that  $\mathbf{\hat{v}}$ and $\mathbf{\hat{n}}$ coincide with $\hat{\mathbf{e}}_{\mu}$, and we recover the standard basic notion. For simplicity in the calculations and interpretation, in most of this work we use adapted coordinates to the geometry of the boundary. 

In this way we can decompose any vector and analyse what happens to its components along tangential part, in this case a surface of dimension 1, that is defined by its normal $\mathbf{\hat{n}}$. This a very involved way of saying we want to study the properties and evolution of the components along $\mathbf{\hat{v}}$ of $\mathbf{T}$, that only depends on the geometry of the problem and not the coordinates. Of course, we can choose adapted coordinates and choose $\mathbf{\hat{v}}$ and $\mathbf{\hat{n}}$ as the $x$ and $y$ axis, and we recover the elemental vector decomposition in cartesian coordinates.

This idea can be extended to tensor fields on $n-$dimensional spacetimes, $\mathcal{ M}$, where a \emph{hypersurface} is an $(n-1)-$dimensional submanifold, $M$, of $\mathcal{ M}$.\footnote{Of course, in the case considered here, $n=3$, $M$ can be simply called surface but we shall keep calling it hypersurface so we can apply directly the discussion here to problems in higher dimensions, for instance $n=4$.}

When the surfaces we want to project on can be either space-like or time-like, we can define a projector, $P$, that projects a tensor field, $T^{\nu}$, into a $(n-1)-$dimensional hypersurface defined by its normal vector, $\mathbf{n}$, as,
\begin{equation}\label{Projector-sigma}
P^{\mu} _{\nu} := \delta^{\mu} _{\nu}  - \sigma n_{\nu} n^{\mu},
\end{equation}
with 
\begin{equation}
\sigma =
\begin{cases}
-1, & \text{if}\ \mathbf{n}\,\, \text{is timelike} \\
1, & \text{if}\ \mathbf{n}\,\, \text{is spacelike}.
\end{cases}
\end{equation}

In particular, we are interested in projecting the fields to a time-like hypersurface (the boundary). To emphasize that we are projecting to a time-like surface which is defined by a space-like normal, that we call $\hat{\mathbf{R}}$ to emphasize this fact. See Figs. (\ref{CSB-Timelike-foliation2}) and (\ref{CSB-Timelike-foliation1}), which shows some possibilities of such time-like surfaces.

\begin{center}
	{	\includegraphics[width=6cm]{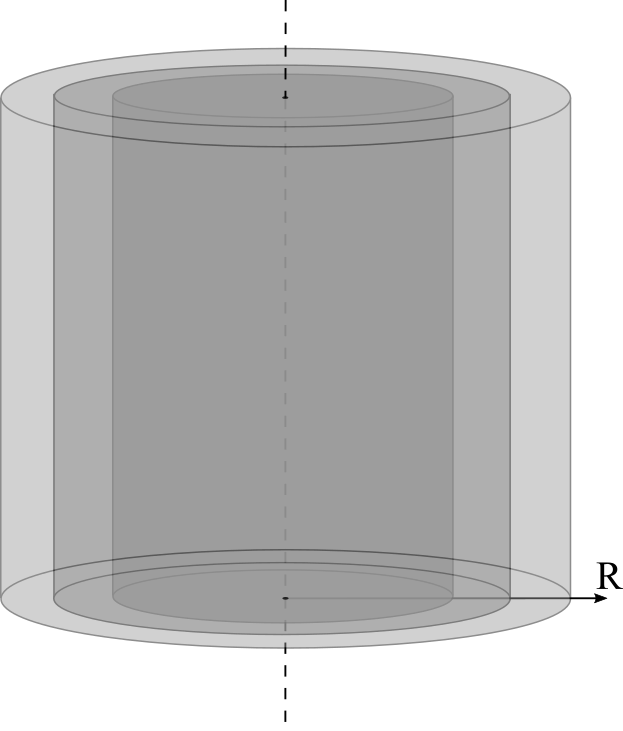}\label{CSB-Timelike-foliation1} \captionof{figure}{Cylindrical foliation: The region $\mathcal{ M}$ of space-time can be foliated by an uniparametric family of cylinders with radius $R$. In this case the normal coincides with the radial direction.} \quad
		\includegraphics[width=6cm]{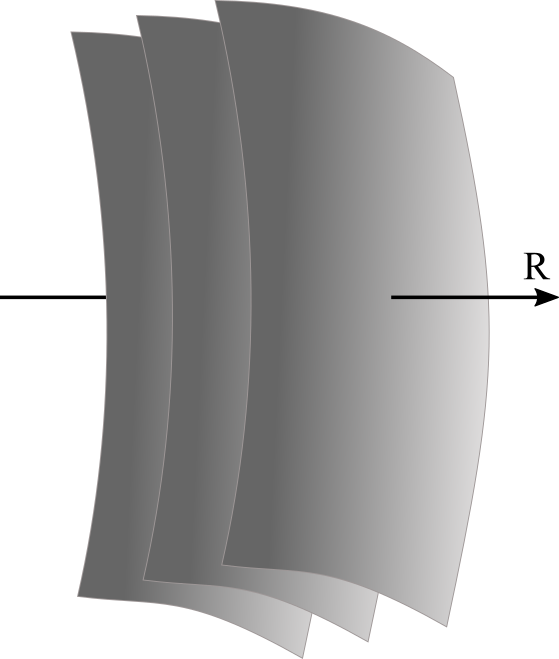}\label{CSB-Timelike-foliation2}}
	\captionof{figure}{The region $\mathcal{ M}$ of space-time can be foliated also by an uniparametric family of open non-intersecting surfaces labeled by $R$. }
\end{center}

Locally these are the two possible foliations, in a topological sense. Other choices, for instance hyperboloids, are topologically equivalent to the cylinders so much of the discussion for the well-posedness of the action depending on one of this choices will be valid for most of physically interesting cases\footnote{We need to be careful since there are subtle differences when using hyperboloids vs cylinders to foliate the space-time in asymptotically flat space-times, see e.g. \cite{Corichi-Reyes-2015}.}.

\subsection{Geometrical definitions}\label{Section-geometry-definitions}

In this section we discuss the technical mathematical aspects of our geometrical construction. Thought we write the paper in such a way that it may be understood even if the details of this section are skipped. Some pedagogical and nice references about hypersurfaces and foliations can be found in \cite{Carrol-2004,Gourgoulhon-2012,Poisson-2004}.

We consider a region of a space-time, $\mathcal{M}$, such that it has topology $\mathbb{R}\times \mathcal{B}$, where $\mathcal{B}$ can be a space-like or time-like hypersurface\footnote{Note that we are not using the term \emph{globally hyperbolic} since it is usually associated with Cauchy slices, spacelike hypersurfaces that need some causal structure, and we are mainly interested in foliating the region of space-time $\mathcal{ M}$ into time-like hypersurfaces. One of this hypersurfaces will become our \emph{boundary}.}. This means that $\mathcal{M}$ can be foliated by a family of hypersurfaces $(\mathcal{B}_{R})_{R\in \mathbb{R}}$. We mean by \emph{foliation} that there exists a smooth scalar field $\mathcal{R}$ on $\mathcal{M}$, which is regular (its gradient never vanishes), such that each hypersurface is a level surface of this scalar field,
\begin{equation}\label{Geometry-Def-foliation}
\forall R \in \mathbb{R}, \,\,\, \mathcal{B}_{R}:=\{ p \in \mathcal{M}, \mathcal{R}(p) = R  \}.
\end{equation}
Two hypersurfaces $\mathcal{B}_{R}$ and $\mathcal{B}_{R'}$ do not intersect,
\begin{equation}
\mathcal{B}_{R}   \cap \mathcal{B}_{R'}  = \emptyset, \,\,\, \text{for} \,\,\,  R \neq R',
\end{equation}
because $\mathcal{R}$ is regular. We call each hypersurface $\mathcal{B}_{R}$ a \emph{leaf} or a \emph{slice} of the foliation. We assume the foliation covers $\mathcal{ M}$,
\begin{equation}
\mathcal{ M} = \bigcup_{R \in \mathbb{R}} \mathcal{ B}_{R}.
\end{equation}

Given the scalar field $\mathcal{R}$ on $\mathcal{ M}$ such that the hypersurface $\mathcal{B}_{R}$ is defined as a level surface of 
$\mathcal{R}$ [cf. Eq. (\ref{Geometry-Def-foliation})], the gradient $1-$form $\mathrm{d} \mathcal{R} $ is normal to $\mathcal{B}_{R}$, such that for every tangent vector $\mathbf{v}$ tangent to $\mathcal{B}_{R}$, $i_{\mathbf{v}}\mathrm{d} \mathcal{R} =0$ (where $i_{\mathbf{v}}\mathrm{d} \mathcal{R}$ denotes the interior product between  $\mathbf{v}$ and $\mathrm{d} \mathcal{R}$, which is also written in other notation as $<\mathrm{d} \mathcal{R},\mathbf{v}> = 0$ ). The vector $\vec{\nabla} \mathcal{R}$ is normal to $\mathcal{B}_{R}$ and satisfy,
\begin{eqnarray}
\bullet & & \vec{\nabla} \mathcal{R} \,\,\,\text{is time-like iff } \,\,\, \mathcal{B}_{R} \,\,\, \text{is space-like},\\
\bullet && \vec{\nabla} \mathcal{R} \,\,\,\text{is space-like iff } \,\,\, \mathcal{B}_{R} \,\,\, \text{is time-like},\\
\bullet && \vec{\nabla} \mathcal{R} \,\,\,\text{is null iff } \,\,\, \mathcal{B}_{R} \,\,\, \text{is null}.
\end{eqnarray}

Note that $\vec{\nabla} \mathcal{R}$ defines a unique direction normal to $\mathcal{B}_{R}$, so any normal vector, $\vec{u}$, must be collinear to $\vec{\nabla} \mathcal{R}$: $\vec{u} = \beta \vec{\nabla} \mathcal{R}$. If $\vec{\nabla} \mathcal{R}$ is not null we can renormalize $\vec{\nabla} \mathcal{R}$ to construct a unit vector,
\begin{equation}
\hat{\mathbf{R}} := \left( \pm \vec{\nabla} \mathcal{R} \cdot \vec{\nabla} \mathcal{R} \right)^{-1/2} \vec{\nabla} \mathcal{R},
\end{equation}
where $+$ is for a time-like hypersurface, while $-$ for a space-like one. And satisfies,
\begin{equation}
\hat{\mathbf{R}} \cdot \hat{\mathbf{R}} =
\begin{cases}
1, & \text{iff}\ \mathcal{B}_{R}\,\, \text{is timelike} \\
-1, & \text{iff}\ \mathcal{B}_{R}\,\, \text{is spacelike}.
\end{cases}
\end{equation}

This construction is not possible for null hypersurfaces because in that case $\vec{\nabla} \mathcal{R} \cdot \vec{\nabla} \mathcal{R}=0$. So, for null hypersurfaces there is no natural way to define a privileged unit normal vector. In fact, given one unit normal vector $\hat{\mathbf{R}} $, another one $\hat{\mathbf{R}} '= \lambda \hat{\mathbf{R}} $ is an equally good alternative. 

\subsubsection{The Orthogonal Projector}

At each point $p \in \mathcal{B}_{R}$, the space of all space-time vector can be orthogonally decomposed as,
\begin{equation}
\mathcal{T}_{p} (\mathcal{ M}) = \mathcal{T}_{p} (\mathcal{ B}_{R}) \oplus \mathrm{span} (\hat{\mathbf{R}}),
\end{equation}
where $	\mathcal{T}_{p} (\mathcal{ M})$ and $\mathcal{T}_{p} (\mathcal{ B}_{R})$ stands out for the tangent spaces to $\mathcal{ M}$ and $\mathcal{ B}_{R}$ at the point $p$ respectively; and $\mathrm{span} (\hat{\mathbf{R}})$ stands for the $1-$dimensional subspace of $	\mathcal{T}_{p} (\mathcal{ M})$ generated by the vector $\hat{\mathbf{R}}$. Which in components with respect any basis  $\hat{\mathbf{e}}_{\mu}$ of $\mathcal{T}_{p} (\mathcal{ M})$ are,
\begin{equation}\label{Projector-R-sigma}
P^{\mu} _{\nu} := \delta^{\mu} _{\nu}  - \sigma R_{\nu} R^{\mu},
\end{equation}
where $\hat{\mathbf{R}} = R^{\mu} \hat{\mathbf{e}}_{\mu}$. With $\sigma = R^{\mu} R_{\mu}$ such that,
\begin{equation}
\sigma =
\begin{cases}
-1, & \text{if}\ \hat{\mathbf{R}}\,\, \text{is timelike} \\
1, & \text{if}\ \hat{\mathbf{R}}\,\, \text{is spacelike}.
\end{cases}
\end{equation}

Note that this orthogonal decomposition holds only for space-like and time-like hypersurfaces, but not for null ones.

We can also use the \emph{projector} $P$ to define an ``orthogonal projection operation'' for all tensors on $\mathcal{M}$ in the following way. We can project the tensor $T^{\alpha_{1} ... \alpha_{p}}\,  _{\beta_{1}...\beta_{q}}$ by applying the projector as many times as indices has $\mathbf{T}$,
\begin{equation}
\underbrace{\mathcal{T}^{\alpha_{1} ... \alpha_{p}} \, _{\beta_{1}...\beta_{q}}}_{Tangential\,\,\,part} := P^{\alpha_{1}} _{\mu_{1}}\cdots P^{\alpha_{p}} _{\mu_{p}} P^{\nu_{1}} _{\beta_{1}} \cdots P^{\nu_{q}} _{\beta_{q}} T^{\mu_{1} ... \mu_{p}}\,  _{\nu_{1}...\nu_{q}}.
\end{equation}

\section{Hamiltonian analysis of the bulk theory in adapted coordinates to the boundary}\label{Appendix-Dirac-bulk-theory}

In this Appendix we shall point out some of the key results for the Chern-Simons bulk theory, but written in the adapted coordinates to the foliation into timelike boundaries labeled by $R$ so we can easily compare with the results in the present work\footnote{You can check the details of the hamiltonian analysis for the non-abelian theory in \cite{Escalante-Carbajal-2011}.}.

For the following analysis we are considering that the region of space-time we are integrating on has topology $\mathcal{M}= \mathbb{R}\times \mathcal{B}$, where $\mathcal{B}$ is a time-like hypersurface. And that each of the hypersurfaces $\mathcal{B}_{R}$ are globally hyperbolic with topology $\Sigma \times \mathbb{R}$, with $\Sigma$ a Cauchy surface.

We start with the projected Chern-Simons action (\ref{CS-Projected-action}), that is equivalent to the Chern-Simons bulk action just decomposed into time-like or space-like surfaces defined by $\hat{\mathbf{R}}$,
\begin{equation}\label{CSBulk-Projected-action}
S_{CS} [A] =  \frac{\kappa}{4} \int_{\mathcal{M}} \left[ (A_{\rho} R^{\rho}) F_{\mu \nu} + 2 A_{\mu} \partial_{\nu} (A_{\rho} R^{\rho}) - 2 A_{\mu} \pounds_{\vec{R}} A_{\nu} \right] \tilde{\varepsilon}^{ \mu \nu} \mathrm{d} R.
\end{equation}

This \emph{projected action} to surfaces, $\mathcal{B}_{R}$, defined by their normal $R^{\mu}$ is still a bulk action, where we just have made evident the separation between quantities that are tangential to the surface and normal. The previous expression holds for any family of surfaces $\mathcal{B}_{R}$ (space-like or time-like) as long as $R^{\mu}$ is the normal to such surfaces.

Note that $(A_{\mu} R^{\mu})$ is a scalar, that we define as $A_{R} := A_{\mu} R^{\mu}$, and $\partial_{R} A_{j} := (\pounds_{\vec{R}} A_{j})$ to simplify notation\footnote{In the case when we chose adapted coordinates to the geometry of the boundary the expressions will take this form. Using this definitions we can recover the most general form of the expressions (independent on the particular choice of coordinates) but also this expressions are ready for doing calculations on adapted coordinates as we show in Sec. \ref{Section-QuatumHallEffect}.}, and whose explicit form can be found once we chose the coordinates. Therefore, in adapted coordinates to the foliation, the Chern-Simons action can be written as,
\begin{equation}\label{Chern-Simons-Bulk-adaptedcoordinates}
S_{CS} [A_{R}, A_{i}] =  \frac{\kappa}{4} \int_{\mathcal{M}} \left[ A_{R}  F_{ij} + 2 A_{i} \partial_{j} A_{R} - 2A_{i} \left( \partial_{R} A_{j}\right)  \right]  \tilde{\varepsilon}^{ij} \mathrm{d} R, 
\end{equation}
were we have changed the labels from greek to latin indices to make clear that the indices now take only the values, $i,j=0,1$. 

Taking the variation of the previous action,
\begin{eqnarray}\label{Variation SCB}
\nonumber \delta S_{CS} [A_{R}, A_{i}] &=&  \frac{\kappa}{2} \int_{\mathcal{M}} \left[ \partial_{i} \left( A_{R} \delta A_{j}- A_{j} \delta A_{R}  \right) - \frac{1}{2} \partial_{R} \left( A_{i} \delta A_{j} - A_{j} \delta A_{i} \right) \right. \\
&& + \left. \left( F_{ij} \right) \delta A_{R} - 2 \left(\partial_{i} A_{R} - \partial_{R} A_{i} \right) \delta A_{j} \right] \tilde{\varepsilon}^{ij} \mathrm{d} R.
\end{eqnarray}
we can obtain the equations of motion, 
\begin{eqnarray}\label{CSB-EqsMotion}
\delta A_{R} &:& \tilde{\varepsilon}^{ij} F_{ij} = 0, \\
\delta A_{j} &:&  \tilde{\varepsilon}^{ij} ( \partial_{i} A_{R} - \partial_{R} A_{i})  = 0.
\end{eqnarray}
A solution, to the previous equations, written in adapted coordinates to the geometry of the boundary $(t,R,\bar{x})$ is,
\begin{equation}
A_{t} = \partial_{t} \phi ,\,\,\,\, A_{\bar{x} } = \partial_{\bar{x}} \phi \,\,\,\,\,  \mathrm{and} \,\,\,\,  A_{R} = \partial_{R} \phi ,
\end{equation}
where $\phi$ is valued in the Lie Algebra of $U(1)$, if the \emph{boundary term},
\begin{equation}\label{CSBulk-Boundaryterm}
\frac{\kappa}{2} \int_{ \mathcal{M}} \left[ \partial_{i} \left( A_{R} \delta A_{j}- A_{j} \delta A_{R}  \right)  - \frac{1}{2} \partial_{R} \left( A_{i} \delta A_{j} - A_{j} \delta A_{i} \right)  \right] \tilde{\varepsilon}^{ij} \mathrm{d} R.
\end{equation}
vanishes under appropriate boundary conditions.

In this Appendix we do not discuss any further about the boundary conditions, since we are just pointing out some key results of the standard bulk analysis, just written in a way that we can easily compare with the main results of this work. The discussion of the well posedness of the action principle is given in Sec. \ref{Section-Analyzing the theory at the boundary}.

\subsection{Hamiltonian analysis}

To be able to count the degrees of freedom of the theory, as well as finding the gauge symmetries, we proceed with a Hamiltonian analysis of the Chern-Simons action (\ref{Chern-Simons-Bulk-adaptedcoordinates}). To do so, we have to decompose the action into the space-like and time-like parts as it is standard \cite{Escalante-Carbajal-2011,Witten-1988a,Witten-1988b}.

Using $\tilde{\varepsilon}^{ij} = 2t^{[i}\tilde{\varepsilon}^{ j]} \mathrm{d} t$,
\begin{eqnarray}\label{Chern-Simons-Bulk-ADM projection0}
\nonumber S_{CS} [A_{R}, A_{i}] &=&  \frac{\kappa}{4} \int_{\mathcal{M}} \left[ A_{R}  F_{ij} + 2 A_{i} \partial_{j} A_{R} - 2A_{i} \partial_{R} A_{j} \right] 2t^{[i}\tilde{\varepsilon}^{ j]} \mathrm{d} t \mathrm{d} R
\\ \nonumber &=&  \frac{\kappa}{4} \int_{\mathcal{M}} \left[ A_{R}  F_{ij} + 2 A_{i} \partial_{j} A_{R} - 2A_{i} \partial_{R} A_{j}  \right] \left[ t^{i}\tilde{\varepsilon}^{ j } - t^{j}\tilde{\varepsilon}^{ i } \right] \mathrm{d} t \mathrm{d} R\\ \nonumber 
&=& \frac{\kappa}{2} \int_{\mathcal{M}} \left[ A_{R} \, t^{i} F_{ij} \,  + (t^{i}A_{i})\partial_{j} A_{R} \,  -  A_{j} t^{i} \partial_{i} A_{R} \,  - t^{i} A_{i} \partial_{R} A_{j} + A_{j} \partial_{R} A_{i} t^{i}  \right] \tilde{\varepsilon}^{j} \mathrm{d} t \mathrm{d} R.\\
\end{eqnarray}
Using that $t^{i} F_{ij} = \pounds_{\vec{t}} A_{j} - \partial_{j} (t^{i}A_{i})$, defining $A_{t} := t^{i}A_{i}$ and remembering we are using adapted coordinates to the foliation,
\begin{equation}
S_{CS} [A_{R}, A_{j}] =  \frac{\kappa}{2} \int_{\mathcal{M}} \left[ A_{R}  \left( \partial_{t} A_{j} - \partial_{j} A_{t} \right)  + A_{t}\partial_{j} A_{R} -  A_{j} \partial_{t} A_{R} - A_{t} \partial_{R} A_{j} +  A_{j} \partial_{R} A_{t}   \right] \tilde{\varepsilon}^{ j} \mathrm{d} t \mathrm{d} R.
\end{equation}
Note that although it seems to be sum over $j$, now $j=1$, so, to avoid confusion in what follows we rename the index $\bar{x} = j$ and the coordinate as $x^{j} := \bar{x}$, and the one-dimensional volume element $\tilde{\varepsilon} := \tilde{\varepsilon}^{ j}$,
\begin{equation}\label{CSBulk-1plus1splitting}
S_{CS} [A_{R}, A_{t}, A_{\bar{x}}] =  \frac{\kappa}{2} \int_{\mathcal{M}} \left[ A_{R}  \partial_{t} A_{\bar{x}} -  A_{\bar{x}} \partial_{t} A_{R}   + A_{t} \partial_{\bar{x}} A_{R}  -  A_{R}  \partial_{\bar{x}} A_{t}  - A_{t} \partial_{R} A_{\bar{x}} +  A_{\bar{x}} \partial_{R} A_{t}   \right] \tilde{\varepsilon} \mathrm{d} t \mathrm{d} R.
\end{equation}
Equation (\ref{CSBulk-1plus1splitting}) \emph{holds for any family of coordinates and for any time-like boundaries}, $\mathcal{B}_{R}$, such that $	\mathcal{ M} = \bigcup_{R \in \mathbb{R}} \mathcal{ B}_{R}$. 
Note that if we want to make explicit calculations in a particular coordinate system with a particular boundary geometry we can express the one-dimensional volume element as $\tilde{\varepsilon} = \sqrt{|\gamma|} d\bar{x}$ with $\gamma$ the metric on the space-like slices of $\mathcal{B}_{R}$. 

In the particular case when we are choosing cylindrical coordinates and cylinders as time-like boundaries, the space-like surfaces are circles $S^{1}$, so $\tilde{\varepsilon} =  \sqrt{|R^{2}|} \mathrm{d} \theta$ with $R=const$ the radius of the cylinder for finite boundaries, but it can be extended to asymptotic boundaries also. Therefore, the previous action takes the form,
\begin{equation}
S_{CS} [A_{R}, A_{i}] =  \frac{\kappa}{2} \int_{\mathcal{M}} \left[ A_{R}  \partial_{t} A_{\theta} -  A_{\theta} \partial_{t} A_{R}   + A_{t} \partial_{\theta} A_{R}  -  A_{R}  \partial_{\theta} A_{t}   -  A_{t} \partial_{R} A_{\theta} +  A_{\theta} \partial_{R} A_{t}  \right] r \mathrm{d} \theta \mathrm{d} t  \mathrm{d} R.
\end{equation}

For the Hamiltonian analysis we find the momenta $(\Pi_{t},\Pi_{R},\Pi_{\bar{x} } )$ canonically conjugated to $(A_{t},A_{R},A_{\bar{x}} )$ are
\begin{eqnarray}
\Pi_{t} &=& \frac{\partial \mathcal{L} }{\partial(\partial_{t} A_{t} )} = 0 \label{Pi-t}
\\	\Pi_{R} &=& \frac{\partial \mathcal{L}  }{\partial(\partial_{t} A_{R} )} = - \alpha A_{\bar{x}}  \label{Pi-R}
\\	\Pi_{\bar{x}} &=& \frac{\partial \mathcal{L} }{\partial(\partial_{t} A_{\bar{x}} )} = \alpha A_{R}   \label{Pi-xbar}
\end{eqnarray}
with $\alpha = (\kappa  \sqrt{|\gamma|})/2 $. Since we can not find all the ``velocities'' with these expressions\footnote{Formally, the determinant of the Hessian matrix is zero, so the system is singular, so we must proceed with the Dirac algorithm.}, we can define the following primary constraints,
\begin{eqnarray}\label{CSB-PrimaryConstraints}
\phi_{t} &:=& \Pi_{t} \approx 0
\\ \phi_{R} &:=&	\Pi_{R}  + \alpha A_{\bar{x}} \approx 0
\\  \phi_{\bar{x}} &:=&	\Pi_{\bar{x}} - \alpha A_{R} \approx 0.
\end{eqnarray}
The Jacobian
\[
\frac{\partial(\phi_{t},\phi_{R},\phi_{\bar{x}})}{\partial(A_{t},A_{R},A_{\bar{x}},\Pi_{t},\Pi_{R},\Pi_{\bar{x}})}=\left(\begin{array}{cccccc}
0 & 0 & 0 & 1 & 0 & 0\\
0 & 0 & \alpha & 0 & 1 & 0 \\
0 & -\alpha & 0 & 0 & 0 & 1
\end{array}\right),
\]
has $\mathrm{Rank} = 3$, which is constant on the constraint surfaces, so these primary constraints (\ref{CSB-PrimaryConstraints}) are a good set of constraints, that is, they define a submanifold on the phase space: the constraint hypersurface.

The local canonical hamiltonian is,
\begin{equation}
H_{C} := \int_{\Sigma} \mathcal{H}_{C} \mathrm{d}\bar{x}  \mathrm{d} R = \int_{\Sigma} \left[ \left( \partial_{t} A_{t} \right) \Pi_{t} + \left( \partial_{t} A_{R} \right) \Pi_{R} + \left( \partial_{t} A_{\bar{x}} \right) \Pi_{\bar{x}}  - \mathcal{L} \right] \mathrm{d}\bar{x}  \mathrm{d} R,
\end{equation}
where $\mathcal{L}$ is the lagrangian density, $\Sigma$ denotes a space-like slice of $\mathcal{M}$ and
\begin{equation}
\mathcal{L} = \alpha  \left[ A_{R}  \partial_{t} A_{\bar{x}} -  A_{\bar{x}} \partial_{t}   + A_{t} \partial_{\bar{x}} A_{R}  -  A_{R}  \partial_{\bar{x}} A_{t}  - A_{t} \partial_{R} A_{\bar{x}} +  A_{\bar{x}} \partial_{R} A_{t}   \right].
\end{equation}
Simplifying,
\begin{equation}
H_{C} := \int_{\Sigma} \mathcal{H}_{C} \mathrm{d}\bar{x}  \mathrm{d} R = - \int_{\Sigma} \alpha  \left[ A_{t} \partial_{\bar{x}} A_{R}  -  A_{R}  \partial_{\bar{x}} A_{t}  - A_{t} \partial_{R} A_{\bar{x}} +  A_{\bar{x}} \partial_{R} A_{t}   \right] \mathrm{d} \bar{x}  \mathrm{d} R.
\end{equation}
We define the \emph{primary hamiltonian} as,
\begin{equation}
H_{1} := H_{C} + \int_{\Sigma} \left[ \lambda_{t} \phi_{t} + \lambda_{R} \phi_{R} + \lambda_{\bar{x}} \phi_{\bar{x}} \right] \mathrm{d}\bar{x}.
\end{equation}
The consistency conditions on the primary constraints are,
\begin{eqnarray}
\nonumber	\dot{\phi }_{t}(x) &:=& \left\{ \phi_{t} (x) , H_{1} \right\} 
= - 2\alpha \,\, \partial_{\bar{x}} A_{R} (x)  + 2 \alpha \partial_{R} A_{\bar{x}} (x)  \approx 0\\
\nonumber	\dot{\phi}_{R}(x) &:=& \left\{ \phi_{R} (x) , H_{1} \right\} 
=  2\alpha \,\, \partial_{\bar{x}} A_{t} (x) + 2 \alpha \lambda_{\bar{x}} (x) \approx 0\\
\nonumber	\dot{\phi}_{\bar{x}}(x) &:=& \left\{ \phi_{\bar{x}} (x) , H_{1} \right\} 
=  - 2\alpha \,\, \lambda_{R} (x) - 2 \partial_{R} A_{t} (x) \approx 0
\end{eqnarray}

From these consistency conditions we have one secondary constraint, $\psi$, and two conditions on the Lagrange multipliers, $\lambda_{\bar{x}}$ and $\lambda_{R}$,
\begin{eqnarray}
\label{CSB-SecondaryConstraint} \psi &:=& - \partial_{\bar{x}} A_{R}  +  \partial_{R} A_{\bar{x}} \approx 0\\
\label{CSBulk-lambdax}	\lambda_{\bar{x}} &=& - \partial_{\bar{x}} A_{t} \\
\label{CSBulk-lambdaR}	\lambda_{R} &=& -  \partial_{R} A_{t}
\end{eqnarray}
Checking consistency condition on the secondary constraint,
\begin{equation}
\nonumber	\dot{\psi}(x) := \left\{ \psi (x) , H_{1} \right\} 
=  \partial_{\bar{x}} \lambda_{R} (x) -  \partial_{R} \lambda_{\bar{x}} (x) \approx 0.
\end{equation}
But notice that taking into account the expressions we found for $\lambda_{R}$ and $\lambda_{\bar{x}}$, equations (\ref{CSBulk-lambdaR}) and (\ref{CSBulk-lambdax}) respectively, and substituting in the previous expression we've got an identity, so we do not have more constraints nor additional information on the Lagrange multipliers. 

So far we have found the following constraints,
\begin{eqnarray}
\phi_{t} &:=& \Pi_{t} \approx 0, \label{phi_t}
\\ \phi_{R} &:=&	\Pi_{R}  + \alpha A_{j} \approx 0,\label{phi_R}
\\  \phi_{\bar{x}} &:=&	\Pi_{\bar{x}} - \alpha A_{R} \approx 0,  \label{phi_x}\\
\psi &:=&  \partial_{\bar{x}} A_{R}  -  \partial_{R} A_{\bar{x}} \approx 0, \label{psi}
\end{eqnarray}
and the conditions on the Lagrange multipliers,
\begin{eqnarray}\label{CSB-LagrangeMultipliers}
\lambda_{\bar{x}} &=& - \partial_{\bar{x}} A_{t} \\
\lambda_{R} &=& -  \partial_{R} A_{t}.
\end{eqnarray}
We need to split the constraints into first and second class. After the separation we have two first class constraints,
\begin{eqnarray}
\gamma_{1} &:=& \Pi_{t} \approx 0\\
\gamma_{2} &:=& \partial_{R} \Pi_{R} + \partial_{\bar{x}} \Pi_{\bar{x}} + 2\alpha \left( \partial_{R} A_{\bar{x}} + \partial_{\bar{x}} A_{R} \right) \approx 0
\end{eqnarray}
and two second class,
\begin{eqnarray}
\chi_{1} &:=& \Pi_{R}  + \alpha A_{j} \approx 0 \\
\chi_{2} &:=& \Pi_{\bar{x}} - \alpha A_{R} \approx 0.
\end{eqnarray}


With this information we are ready to count the degrees of freedom on the bulk,
\begin{eqnarray}
\nonumber D.O.F. &=& \frac{1}{2} \left[(\#phase\,\, space \,\, variables) - 2 (\# 1st-class) - (\#2nd-class)\right]\\
&=& \frac{1}{2} \left[(6) - 2 (2) - (2)\right] = \mathbf{0}, \label{DOF-CS-bulk}
\end{eqnarray}
which is the well known fact that the $3-$dimensional Chern-Simons theory is topological.


We know that the first class constraints are generators of the gauge transformations. Given a dynamical variable $F(q,p)$ we can calculate its gauge transformations by,
\begin{equation}
\delta F (x) = \int \varepsilon^{\alpha} (y) \{ F(x) , \gamma_{\alpha} (y) \} \mathrm{d}y,
\end{equation}
where the index $\alpha$ runs over all the first class constraints, $\gamma_{\alpha}$, and $\varepsilon^{\alpha}$ is the infinitesimal gauge parameter. 
Therefore, the gauge transformations are,
\begin{eqnarray}\label{CS-Boundary-GeneralGaugeTransformations}
\nonumber	\delta A_{t} (x) &= & \partial_{t} \varepsilon (x),\\
\delta A_{\bar{x}} (x) &=& \partial_{\bar{x}} \varepsilon (x),\\
\delta A_{R} (x) &=& \partial_{R} \varepsilon (x).
\end{eqnarray}
Which we can write in the standard way as,
\begin{equation}
A'_{\mu} = A_{\mu} + \partial_{\mu} \varepsilon
\end{equation}
with $\mu = t,\bar{x},R$. Choosing $\varepsilon = v^{\alpha} A_{\alpha}$, we can also see that the diffeomorphisms are a symmetry of the theory,
\begin{equation}
A'_{\mu} = A_{\mu} + v^{\alpha} F_{\mu \alpha} + \pounds_{\vec{v}} A_{\mu}.
\end{equation}


\section{Hamiltonian analysis of the complete boundary theory }\label{Hamiltonian-analysis-complete-theory}

To be able to count the degrees of freedom of the theory in the boundary as well as finding the gauge symmetries on the boundary we shall proceed with a Hamiltonian analysis of the action (\ref{Chern-Simons-Boundary}), that is the Chern-Simons  theory at the boundary,
\begin{equation}
S_{CSB} [A_{R}, A_{i}, \lambda] =  \frac{\kappa}{4} \int_{\mathcal{M}} \left[ A_{R}  F_{ij} + 2 A_{i} \partial_{j} A_{R} - 2 (1 + \lambda) A_{i} \left( \partial_{R} A_{j}\right)  \right]  \tilde{\varepsilon}^{ij} \mathrm{d} R.
\end{equation}

To proceed with the Hamiltonian analysis we have to decompose the action into the space-like and time-like parts as it is standard (see for instance, \cite{Romano-1993}). Using $\tilde{\varepsilon}^{ij} = 2t^{[i}\tilde{\varepsilon}^{ j]} \mathrm{d} t$
\begin{eqnarray}\label{Chern-Simons-Boundary-ADM projection0}
S_{CSB} [A_{R}, A_{j}, \lambda] &=&  \frac{\kappa}{4} \int_{\mathcal{M}} \left[ A_{R}  F_{ij} + 2 A_{i} \partial_{j} A_{R} - 2 (1+ \lambda) A_{i} \partial_{R} A_{j} \right] 2t^{[i}\tilde{\varepsilon}^{ j]} \mathrm{d} t
\\ &=&  \frac{\kappa}{4} \int_{\mathcal{M}} \left[ A_{R}  F_{ij} + 2 A_{i} \partial_{j} A_{R} - 2 (1+ \lambda) A_{i} \partial_{R} A_{j}  \right] \left[ t^{i}\tilde{\varepsilon}^{ j } - t^{j}\tilde{\varepsilon}^{ i } \right] \mathrm{d} t\\
\nonumber &=& \frac{\kappa}{2} \int_{\mathcal{M}} \left[ A_{R} \, t^{i} F_{ij} \,  + (t^{i}A_{i})\partial_{j} A_{R} \,  -  A_{j} t^{i} \partial_{i} A_{R} \,   \right.\\
&& \left.  - (1+ \lambda) t^{i} A_{i} \partial_{R} A_{j} + (1+ \lambda) A_{j} \partial_{R} A_{i} t^{i}  \right] \tilde{\varepsilon}^{j} \mathrm{d} t
\end{eqnarray}
using that $t^{i} F_{ij} = \pounds_{\vec{t}} A_{j} - \partial_{j} (t^{i}A_{i})$, defining $A_{t} := t^{i}A_{i}$ to simplify notation, and  then using adapted coordinates, $\pounds_{\vec{t}} = \partial_{t}$
\begin{eqnarray}
\nonumber S_{CSB} [A_{R}, A_{j}, \lambda] &=&  \frac{\kappa}{2} \int_{\mathcal{M}} \left[ A_{R}  \left(  \partial_{t} A_{j} - \partial_{j} A_{t} \right)  + A_{t}\partial_{j} A_{R} -  A_{j}  \partial_{t} A_{R} \right. \\
&& \left. - (1+ \lambda) A_{t} \partial_{R} A_{j} + (1+ \lambda)  A_{j} \partial_{R} A_{t}   \right] \tilde{\varepsilon}^{ j} \mathrm{d} t.
\end{eqnarray}
Note that although it seems to be sum over $j$, now $j=1$, so, to avoid confusion in what follows we rename the index $\bar{x} = j$ and the coordinate as $x^{j} := \bar{x}$, and the one-dimensional volume element $\tilde{\varepsilon} := \tilde{\varepsilon}^{ j}$. So we can write the \emph{complete action principle} that contains the information of the dynamics in each time-like hypersuface as,
\begin{eqnarray}\label{CSB-complete-action-1plus1splitting}
\nonumber S_{CSB} [A_{R}, A_{t}, A_{\bar{x}}, \lambda] &=&  \frac{\kappa}{2} \int_{\mathcal{M}} \left[ A_{R}  \partial_{t} A_{\bar{x}} -  A_{\bar{x}} \partial_{t} A_{R}   + A_{t} \partial_{\bar{x}} A_{R}  -  A_{R}  \partial_{\bar{x}} A_{t} \right. \\
&& \left.  - (1+ \lambda) A_{t} \partial_{R} A_{\bar{x}} + (1+ \lambda)  A_{\bar{x}} \partial_{R} A_{t}   \right] \tilde{\varepsilon} \mathrm{d} t.
\end{eqnarray}
Equation (\ref{CSB-complete-action-1plus1splitting}) \emph{holds for any coordinates and for any time-like boundary} $\mathcal{B}$. 
Note that if we want to make explicit calculations in a particular coordinate system with a particular boundary geometry we can express the one-dimensional volume element as $\tilde{\varepsilon} = \sqrt{|\gamma|} d\bar{x}$ with $\gamma$ the metric on the space-like slices of $\mathcal{B}$. 

In the particular case when we are choosing cylindrical coordinates and cylinders as time-like boundaries, the space-like surfaces are circles $S^{1}$, so $\tilde{\varepsilon} =  \sqrt{|R^{2}|} \mathrm{d} \theta$ with $R=const$ the radius of the cylinder for finite boundaries, but it can be extended to asymptotic boundaries also (work in progress). Therefore, the previous action takes the form,
\begin{eqnarray}
\nonumber S_{CSB} [A_{R}, A_{j}, \lambda] &=&  \frac{\kappa}{2} \int_{\mathcal{M}} \left[ A_{R}  \partial_{t} A_{\theta} -  A_{\theta} \partial_{t} A_{R}   + A_{t} \partial_{\theta} A_{R}  -  A_{R}  \partial_{\theta} A_{t} \right.\\
&& \left.  - (1+ \lambda) A_{t} \partial_{R} A_{\theta} + (1+ \lambda) A_{\theta} \partial_{R} A_{t}  \right] r \mathrm{d} \theta \mathrm{d} t.
\end{eqnarray}

In what follows we use equation (\ref{CSB-complete-action-1plus1splitting}) that is independent of the particular choice of coordinates and the geometry of the time-like boundary, so our results are as general as possible considering the minimal requirements we have made so far: choosing the boundary to be time-like, and a direction for time-evolution.

For the Hamiltonian analysis we find that the momenta $(\Pi_{t},\Pi_{R},\Pi_{\bar{x} },\Pi_{\lambda} )$ canonically conjugated to $(A_{t},A_{R},A_{\bar{x}},\lambda )$ are
\begin{eqnarray}
\Pi_{t} &=& \frac{\partial \mathcal{L} }{\partial(\partial_{t} A_{t} )} = 0, \label{Pi-t}
\\	\Pi_{R} &=& \frac{\partial \mathcal{L}  }{\partial(\partial_{t} A_{R} )} = - \alpha A_{\bar{x}},  \label{Pi-R}
\\	\Pi_{\bar{x}} &=& \frac{\partial \mathcal{L} }{\partial(\partial_{t} A_{\bar{x}} )} = \alpha A_{R} ,  \label{Pi-xbar}
\\  \Pi_{\lambda} &=& \frac{\partial \mathcal{L} }{\partial(\partial_{t} \lambda  )} = 0, \label{Pi-lambda}
\end{eqnarray}
with $\alpha = (\kappa  \sqrt{|\gamma|})/2 $. Since we cannot find all the velocities with these expressions\footnote{Formally, the determinant of the Hessian matrix is zero, so the system is singular, so we must proceed with the Dirac algorithm.}, we can define the following primary constraints,
\begin{eqnarray}\label{CSB-PrimaryConstraints}
\phi_{t} &:=& \Pi_{t} \approx 0,
\\ \phi_{R} &:=&	\Pi_{R}  + \alpha A_{\bar{x}} \approx 0,
\\  \phi_{\bar{x}} &:=&	\Pi_{\bar{x}} - \alpha A_{R} \approx 0,
\\  \phi_{\lambda} &:=& \Pi_{\lambda} \approx 0.
\end{eqnarray}
The Jacobian
\[
\frac{\partial(\phi_{t},\phi_{R},\phi_{\bar{x}},\phi_{\lambda})}{\partial(A_{t},A_{R},A_{\bar{x}},\lambda,\Pi_{t},\Pi_{R},\Pi_{\bar{x}},\Pi_{\lambda})}=\left(\begin{array}{cccccccc}
0 & 0 & 0 & 0 & 1 & 0 & 0 & 0\\
0 & 0 & \alpha & 0 & 0 & 1 & 0 & 0\\
0 & -\alpha & 0 & 0 & 0 & 0 & 1 & 0\\
0 & 0 & 0 & 0 & 0 & 0 & 0 & 1
\end{array}\right),
\]
has $\mathrm{Rank} = 4$, which is constant on the constraint surfaces, so these primary constraints (\ref{CSB-PrimaryConstraints}) form, in fact, a submanifold of the phase space.

The local canonical hamiltonian is
\begin{equation}
H_{C} := \int_{\Sigma} \mathcal{H}_{C} \mathrm{d}\bar{x} = \int_{\Sigma} \left[ \left( \partial_{t} A_{t} \right) \Pi_{t} + \left( \partial_{t} A_{R} \right) \Pi_{R} + \left( \partial_{t} A_{\bar{x}} \right) \Pi_{\bar{x}} + \left( \partial_{t} \lambda \right) \Pi_{\lambda}  - \mathcal{L} \right] \mathrm{d}\bar{x},
\end{equation}
where $\mathcal{L}$ is the lagrangian density, $\Sigma$ denotes a space-like slice of $\mathcal{M}$ and
\begin{equation}
\mathcal{L} = \alpha  \left[ A_{R}  \partial_{t} A_{\bar{x}} -  A_{\bar{x}} \partial_{t} A_{R}   + A_{t} \partial_{\bar{x}} A_{R}  -  A_{R}  \partial_{\bar{x}} A_{t}  - (1 + \lambda)  A_{t} \partial_{R} A_{\bar{x}} +  (1 + \lambda) A_{\bar{x}} \partial_{R} A_{t}   \right].
\end{equation}
Simplifying,
\begin{equation}
H_{C} := \int_{\Sigma} \mathcal{H}_{C} \mathrm{d}\bar{x} = - \int_{\Sigma} \alpha  \left[ A_{t} \partial_{\bar{x}} A_{R}  -  A_{R}  \partial_{\bar{x}} A_{t}  - (1 + \lambda) A_{t} \partial_{R} A_{\bar{x}} + (1 + \lambda)  A_{\bar{x}} \partial_{R} A_{t}   \right] \mathrm{d} \bar{x}.
\end{equation}
We define the \emph{primary hamiltonian} as,
\begin{equation}
H_{1} := H_{C} + \int_{\Sigma} \left[ \lambda_{t} \phi_{t} + \lambda_{R} \phi_{R} + \lambda_{\bar{x}} \phi_{\bar{x}} + \lambda_{\lambda} \phi_{\lambda} \right] \mathrm{d}\bar{x}.
\end{equation}
The consistency conditions on the primary constraints are,
\begin{eqnarray}
\nonumber	\dot{\phi }_{t}(\bar{x}) &:=& \left\{ \phi_{t} (\bar{x}) , H_{1} \right\} 
= 2\alpha \,\, \partial_{\bar{x}} A_{R} (\bar{x})  - 2  \alpha (1 + \lambda) \partial_{R} A_{\bar{x}} (\bar{x})  - A_{\bar{x}} \partial_{R} \lambda   \approx 0,\\
\nonumber	\dot{\phi}_{R}(x) &:=& \left\{ \phi_{R} (x) , H_{1} \right\} 
=   -  2\alpha \,\, \partial_{\bar{x}} A_{t} (x) + 2 \alpha \lambda_{\bar{x}} (x) \approx 0,\\
\nonumber	\dot{\phi}_{\bar{x}}(x) &:=& \left\{ \phi_{\bar{x}} (x) , H_{1} \right\} 
=  - 2\alpha \,\, \lambda_{R} (x) + 2 \alpha (1+\lambda) \partial_{R} A_{t} (x) + \alpha A_{t} \partial_{R} \lambda \approx 0,\\
\nonumber	\dot{\phi}_{\lambda}(x) &:=& \left\{ \phi_{\lambda} (x) , H_{1} \right\} 
= - \alpha \left( A_{t} \partial_{R} A_{\bar{x}} - A_{\bar{x}} \partial_{R} A_{t} \right) \approx 0.
\end{eqnarray}

From the consistency conditions on the primary constraints we have found two secondary constraints, $\psi_{t}$ and $\psi_{\lambda}$, and two conditions on the Lagrange multipliers, $\lambda_{\bar{x}}$ and $\lambda_{R}$,
\begin{eqnarray}
\label{CSB-SecondaryConstraint-Psi_t} \psi_{t} &:=&  2 \alpha \partial_{\bar{x}} A_{R}  - 2\alpha (1 + \lambda)  \partial_{R} A_{\bar{x}} - A_{\bar{x}} \partial_{R} \lambda  \approx 0,\\
\label{CSB-SecondaryConstraint-Psi_lambda} \psi_{\lambda} &:=&  - \alpha \left( A_{t} \partial_{R} A_{\bar{x}} - A_{\bar{x}} \partial_{R} A_{t} \right) \approx 0,\\
\label{CSB-lambdax}	\lambda_{\bar{x}} &=&  \partial_{\bar{x}} A_{t}, \\
\label{CSB-lambdaR}	\lambda_{R} &=&   (1 + \lambda)  \partial_{R} A_{t} + \frac{1}{2} A_{t} \partial_{R} \lambda.
\end{eqnarray}
Checking consistency condition on the secondary constraints,
\begin{eqnarray}
\nonumber	\dot{\psi}_{t}(x) &:=& \left\{ \psi (x) , H_{1} \right\} 
=  2 \alpha  \partial_{\bar{x}} \lambda_{R} (x) -  2 \alpha (1 + \lambda)\partial_{R} \lambda_{\bar{x}} (x) - 2 \alpha \lambda_{\lambda} (\bar{x}) \partial_{R} A_{\bar{x}} (x) \approx 0.
\end{eqnarray}
But notice that taking into account the expressions we found for $\lambda_{R}$ and $\lambda_{\bar{x}}$, equations (\ref{CSB-lambdaR}) and (\ref{CSB-lambdax}) respectively, and substituting in the previous expression we find an expression for the Lagrange multiplier $\lambda_{\lambda}$,
\begin{equation}
\lambda_{\lambda} \approx (\partial_{\bar{x}} \lambda)(\partial_{R} A_{t})(\partial_{R} A_{\bar{x}})^{-1}.
\end{equation}
and,
\begin{eqnarray}
\nonumber	\dot{\psi}_{\lambda}(x) &:=& \left\{ \psi (x) , H_{1} \right\} 
=  2 \alpha \lambda_{t} (x) \partial_{R} A_{\bar{x}} (x) - 2 \alpha \lambda_{\bar{x}} (x) \partial_{R} A_{t} (x)  \approx 0,
\end{eqnarray}
But taking into account the expression $\lambda_{\bar{x}}$, we find an expression for the Lagrange multiplier $\lambda_{t}$,
\begin{equation}
\lambda_{t} \approx (\partial_{\bar{x}} A_{t})(\partial_{R} A_{t})(\partial_{R} A_{\bar{x}})^{-1}.
\end{equation}

Note that from the consistency conditions on the secondary constraints we do not get more constraints, only conditions on the Lagrange multipliers. So, we have found the following constraints,
\begin{eqnarray}
\phi_{t} &:=& \Pi_{t} \approx 0, \label{phi_t}
\\ \phi_{R} &:=&	\Pi_{R}  + \alpha A_{\bar{x}} \approx 0,\label{phi_R}
\\  \phi_{\bar{x}} &:=&	\Pi_{\bar{x}} - \alpha A_{R} \approx 0,  \label{phi_x}
\\  \phi_{\lambda} &:=& \Pi_{\lambda} \approx 0, \label{phi_lambda}
\\ \psi_{t} &:=&  2 \alpha \partial_{\bar{x}} A_{R}  - 2\alpha (1 + \lambda)  \partial_{R} A_{\bar{x}} - \alpha A_{\bar{x}} \partial_{R} \lambda \approx 0, \label{psi_t}
\\ \psi_{\lambda} &:=& - \alpha \left( A_{t} \partial_{R} A_{\bar{x}} - A_{\bar{x}} \partial_{R} A_{t} \right) \approx 0, \label{psi_lambda}
\end{eqnarray}
and the conditions on the Lagrange multipliers,
\begin{eqnarray}\label{CSB-LagrangeMultipliers}
\nonumber	\lambda_{\bar{x}} &=&  \partial_{\bar{x}} A_{t}, 
\\  \nonumber	\lambda_{R} &=&   (1 + \lambda)  \partial_{R} A_{t} + \frac{1}{2} A_{t} \partial_{R} \lambda,
\\ \nonumber 	\lambda_{\lambda} &=& (\partial_{\bar{x}} \lambda)(\partial_{R} A_{t})(\partial_{R} A_{\bar{x}})^{-1},
\\ \lambda_{t} &=& (\partial_{\bar{x}} A_{t})(\partial_{R} A_{t})(\partial_{R} A_{\bar{x}})^{-1}.
\end{eqnarray}

Now, after finding all the constraints, we have to separate them into first class and second class. For doing so, we need to calculate the matrix $W$ whose entries are the Poisson brackets among all the primary and secondary constraints,

{\small
	\[
	\hspace{-.3cm}	W =\left(\begin{array}{cccccc}
	\left\{ \phi_{t} (x) , \phi_{t} (y)  \right\}  & 	\left\{ \phi_{t} (x) , \phi_{R} (y) \right\}  & 	\left\{ \phi_{t} (x) , \phi_{\bar{x}} (y) \right\}  & 	\left\{ \phi_{t} (x) , \phi_{\lambda} (y)  \right\}  & 	\left\{ \phi_{t} (x) ,  \psi_{t} (y)  \right\}  & 	\left\{ \phi_{t} (x) , \psi_{\lambda} (y)  \right\}  \\
	
	\left\{ \phi_{R} (x)  , \phi_{t} (y) \right\} & \left\{ \phi_{R} (x)  , \phi_{R} (y) \right\} &  \left\{ \phi_{R} (x)  , \phi_{\bar{x}} (y)  \right\} &  \left\{ \phi_{R} (x)  , \phi_{\lambda} (y) \right\} &  \left\{ \phi_{R} (x)  ,  \psi_{t} (y)   \right\} &  \left\{ \phi_{R} (x)  , \psi_{\lambda} (y) \right\} \\
	
	\left\{ \phi_{\bar{x}} (x)  , \phi_{t} (y)  \right\} & 	\left\{ \phi_{\bar{x}} (x)  , \phi_{R} (y) \right\} &  	\left\{ \phi_{\bar{x}} (x)  , \phi_{\bar{x}} (y) \right\} &  	\left\{ \phi_{\bar{x}} (x)  , \phi_{\lambda} (y) \right\} &  	\left\{ \phi_{\bar{x}} (x)  ,  \psi_{t} (y)  \right\} & 	\left\{ \phi_{\bar{x}} (x)  , \psi_{\lambda} (y)  \right\} \\
	
	\left\{ \phi_{\lambda} (x)  , \phi_{t} (y)  \right\} &	\left\{ \phi_{\lambda} (x)  , \phi_{R} (y) \right\} & 	\left\{ \phi_{\lambda} (x)  , \phi_{\bar{x}} (y) \right\} &	\left\{ \phi_{\lambda} (x)  , \phi_{\lambda} (y) \right\} &	\left\{ \phi_{\lambda} (x)  ,  \psi_{t} (y)  \right\} &	\left\{ \phi_{\lambda} (x)  , \psi_{\lambda} (y) \right\} \\
	
	\left\{ \psi_{t} (x)  , \phi_{t} (y)  \right\} &	\left\{ \psi_{t} (x)  , \phi_{R} (y) \right\} &	\left\{ \psi_{t} (x)  , \phi_{\bar{x}} (y) \right\} &	\left\{ \psi_{t} (x)  , \phi_{\lambda} (y)  \right\} &	\left\{ \psi_{t} (x)  ,  \psi_{t} (y)  \right\} &	\left\{ \psi_{t} (x)  , \psi_{\lambda} (y) \right\} \\
	
	\left\{ \psi_{\lambda} (x)  , \phi_{t} (y)  \right\} &	\left\{ \psi_{\lambda} (x)  , \phi_{R} (y) \right\} &	\left\{ \psi_{\lambda} (x)  , \phi_{\bar{x}} (y) \right\} &	\left\{ \psi_{\lambda} (x)  , \phi_{\lambda} (y) \right\} &	\left\{ \psi_{\lambda} (x)  ,  \psi_{t} (y)  \right\} &	\left\{ \psi_{\lambda} (x)  , \psi_{\lambda} (y) \right\} \\
	
	\end{array}\right)
	\]
}{\small \par}

{\small
	\begin{equation} \label{MatrixW}
	\hspace{-.5cm}	= 2\alpha \left(\begin{array}{cccccc}
	0 & 	0  &  0  & 	0  & 	0  & 	+	 \partial_{R} A_{\bar{x}}  \\
	
	0 & 0 & 1 &  0 &  - \partial_{\bar{x}} & 0 \\
	
	0 & 	-1 &  0 &  	0 &   (1 + \lambda) \partial_{R} + \frac{1}{2} \partial_{R} \lambda & 	- \frac{1}{2}( \partial_{R} A_{t} + A_{t} \partial_{R}) \\
	
	0 &	0 & 	0 &	0 &	  \partial_{R} A_{\bar{x}} + \frac{1}{2} A_{\bar{x}} \partial_{R}  &	0 \\
	
	0 &	 \partial_{\bar{x}} &	- (1 + \lambda) \partial_{R} - \frac{1}{2} \partial_{R} \lambda &	-  \partial_{R} A_{\bar{x}} -\frac{1}{2} A_{\bar{x}} \partial_{R} &	0 &	0  \\
	
	- \partial_{R} A_{\bar{x}} &	0 &	 \frac{1}{2}( \partial_{R} A_{t} + A_{t} \partial_{R}) &	 0 &	0 &	0 \\
	
	\end{array}\right) \delta (x-y)
	\end{equation}
}{\small \par}

The $\mathrm{Rank(W)}=6$, therefore the theory only have the following \emph{second class contraints},
\begin{eqnarray}
\chi_{1} &:=& \phi_{t} = \Pi_{t} \approx 0, \label{chi_1}
\\ \chi_{2} &:=& \phi_{R} =	\Pi_{R}  + \alpha A_{\bar{x}} \approx 0,\label{chi_2}
\\ \chi_{3} &:=& \phi_{\bar{x}} =	\Pi_{\bar{x}} - \alpha A_{R} \approx 0,  \label{chi_3}
\\ \chi_{4} &:=& \phi_{\lambda} = \Pi_{\lambda} \approx 0 \label{chi_4}
\\ \chi_{5} &:=& \psi_{t} =  2 \alpha \partial_{\bar{x}} A_{R}  - 2\alpha (1 + \lambda)  \partial_{R} A_{\bar{x}} - \alpha A_{\bar{x}} \partial_{R} \lambda \approx 0 \label{chi_5}
\\ \chi_{6} &:=& \psi_{\lambda} = - \alpha \left( A_{t} \partial_{R} A_{\bar{x}} - A_{\bar{x}} \partial_{R} A_{t} \right) \approx 0 \label{chi_6}
\end{eqnarray}

Note the difference with the \emph{bulk theory}, where we have both \emph{first class} and \emph{second class constraints}, so it is a gauge theory. Their gauge symmetries are the three-dimensional diffeomorphisms (See Appendix \ref{Appendix-Dirac-bulk-theory} for a summary of the main results in the bulk theory and \cite{Escalante-Carbajal-2011} for the details of the calculation.) In this case we only have second class constraints, so \emph{the theory at the boundary is not a gauge theory.} 


With this information we are ready to count the \emph{degrees of freedom} per slice of the time-like foliation, 
\begin{eqnarray}
\nonumber D.O.F. &=& \frac{1}{2} \left[(\#phase\,\, space \,\, variables) - 2 (\# 1st-class) - (\#2nd-class)\right]\\
&=& \frac{1}{2} \left[(8) - 2 (0) - (6)\right] = \mathbf{1}. \label{DOF-CS-boundary}
\end{eqnarray}

Note one key difference from the theory in the bulk, while the three-dimensional abelian Chern-Simons theory is a topological theory (see e.g. Eq. (\ref{DOF-CS-bulk})), the boundary theory has one degree of freedom. This shows that a theory that has zero local degrees of freedom in the bulk may be non-trivial at the boundary. This has been observed previously for the Chern-Simons theory through different approaches and assumptions. For instance, if you make a partial gauge fixing in the bulk, introducing the solution of \emph{the condition of the restriction to the boundary} as a gauge condition, this partially fixes the gauge freedom, and that  will introduce a degree of freedom coming from the partial breaking of the gauge symmetry. We are not taking that approach here, from first principles and without gauge fixings in the bulk we arise at the same conclusion, that the solution of \emph{the condition of the restriction to the boundary} is a second class constraint and that the theory in the boundary is no longer a gauge theory.


Now that we have all the information of the system we can write the extended hamiltonian and actions. By definition the extended hamiltonian is,
\begin{equation}
H_{E} = \int_{\Sigma} \mathcal{H}_{E} \,\, d\bar{x} := H_{C} + \int_{\Sigma} \left( u^{1} \chi_{1} + u^{2} \chi_{2} + u^{3} \chi_{3} + u^{4} \chi_{4}+  u^{5} \chi_{5}  +u^{6} \chi_{6}  \right) \mathrm{d}\bar{x}\mathrm{d}R,
\end{equation}
for the Chern-Simons theory on the boundary this becomes,
\begin{eqnarray}
\nonumber	H_{E} &=& \int_{\Sigma}  \left[  \alpha A_{R}  \partial_{\bar{x}} A_{t} -  \alpha A_{t} \partial_{\bar{x}} A_{R}  +  \alpha (1 + \lambda) \left( A_{t} \partial_{R} A_{\bar{x}} -    A_{\bar{x}} \partial_{R} A_{t} \right) \right.\\
&& \left. + u^{1} \chi_{1} + u^{2} \chi_{2} + u^{3} \chi_{3} + u^{4} \chi_{4}+  u^{5} \chi_{5}  +u^{6} \chi_{6}   \right]\mathrm{d}\bar{x}\mathrm{d}R.
\end{eqnarray}
Solving the second class constraints, the reduced extended hamiltonian becomes,
\begin{equation}
H_{E-Reduced} = \int_{\Sigma}  \left[   \alpha A_{R}  \partial_{\bar{x}} A_{t} -  \alpha A_{t} \partial_{\bar{x}} A_{R}   \right]\mathrm{d}\bar{x}\mathrm{d}R.
\end{equation}
Note that the term $  \alpha (1 + \lambda)( A_{t} \partial_{R} A_{\bar{x}} -  A_{\bar{x}} \partial_{R} A_{t} )$ vanishes when solving \emph{the condition of the restriction to the boundary}, whose general solution is $A_{t} (t,R,\bar{x}) - v(t,\bar{x}) A_{\bar{x}}(t,R,\bar{x})= 0$, changing  $A_{t} (t,R,\bar{x})$ by $ v(t,\bar{x}) A_{\bar{x}}(t,R,\bar{x})$ in the extended action we have,
\begin{equation}
H_{E-Reduced} = \int_{\Sigma}  \left[   \alpha v(t,\bar{x}) \left(   A_{R}  \partial_{\bar{x}} A_{\bar{x}} -  A_{\bar{x}} \partial_{\bar{x}} A_{R} \right) + \alpha A_{R} A_{\bar{x}} \partial_{\bar{x}} v(t,\bar{x})    \right]\mathrm{d}\bar{x}\mathrm{d}R.
\end{equation}
Then, the extended action is,
\begin{eqnarray}
S_{E-Reduced}[A_{\mu}, \Pi_{\mu}] &:=& \int_{\mathcal{M}} \left[ \Pi_{\mu} \partial_{t} {A}_{\mu}  - \mathcal{H}_{E} \right]\mathrm{d}\bar{x}\mathrm{d}R \mathrm{d}t\\
\nonumber   &=&	\int_{\mathcal{M}} \left[ \Pi_{R} \partial_{t} {A}_{R}  + \Pi_{\bar{x}} \partial_{t} {A}_{\bar{x}}   \right.\\
\nonumber   && \left. -  \alpha v(t,\bar{x}) \left(   A_{R}  \partial_{\bar{x}} A_{\bar{x}} -  A_{\bar{x}} \partial_{\bar{x}} A_{R} \right) - \alpha A_{R} A_{\bar{x}} \partial_{\bar{x}} v(t,\bar{x})    \right] \mathrm{d}\bar{x} \mathrm{d}R \mathrm{d}t.
\end{eqnarray}
Solving $ \chi_{2} := \phi_{R} =	\Pi_{R}  + \alpha A_{\bar{x}} \approx 0$ and $\chi_{3} := \phi_{\bar{x}} =	\Pi_{\bar{x}} - \alpha A_{R} \approx 0$,
\begin{equation}\label{CSB-Reduced-extended-action}
S_{E-Reduced}[A_{\mu}, \Pi_{\mu}]= \int_{\mathcal{M}} \left[  \Pi_{\bar{x}} \left( \partial_{t} A_{\bar{x}} -  \partial_{\bar{x}} \left( v(t,\bar{x}) A_{\bar{x}} \right)  \right) - A_{\bar{x}} \left( \partial_{t} \Pi_{\bar{x}} - v(t,\bar{x}) \partial_{\bar{x}} \Pi_{\bar{x}} \right)    \right] \mathrm{d}\bar{x} \mathrm{d}R \mathrm{d}t.
\end{equation}
whose equations of motion are,
\begin{eqnarray}\label{CSB-Reduced-equations-motion-from-SERreduced}
\partial_{t} A_{\bar{x}} -  \partial_{\bar{x}} \left[ v(t,\bar{x}) A_{\bar{x}} \right] &=& 0\\
\partial_{t} \Pi_{\bar{x}} - v(t,\bar{x}) \partial_{\bar{x}} \Pi_{\bar{x}} &=& 0,
\end{eqnarray}
or equivalently,
\begin{eqnarray}\label{CSB-Reduced-equations-motion-from-SERreduced-intermsof-A}
\partial_{t} A_{\bar{x}} -  \partial_{\bar{x}} \left[ v(t,\bar{x}) A_{\bar{x}} \right] &=& 0\\
\partial_{t} A_{R} - v(t,\bar{x}) \partial_{\bar{x}} A_{R} &=& 0.
\end{eqnarray}
This equations encode the general dynamics of the theory at the boundary, that is, all the possible solutions compatible with the boundary conditions such that we have a well-posed action principle. For instance, considering the boundary conditions, $A_{\bar{x}} |_{t_{1},x_{1}}  = w  A_{R}|_{t_{1},x_{1}} \,\,\, with \,\,\, w = const$, in section \ref{Subsection-Well-posedness-action-Boundary-conditions}, we can see that from \ref{CSB-Reduced-equations-motion-from-SERreduced-intermsof-A} and  the unicity and existence theorem that we have only one equation of motion satisfying the boundary (initial) conditions,
\begin{equation}
\partial_{t} A_{\bar{x}} - v \partial_{\bar{x}} A_{\bar{x} =0,}
\end{equation}
with $v = const.$ That is the equation for a chiral boson. 

\subsection{Remarks on the Hamiltonian analysis of the complete action}\label{Dirac-reduced-system}

It is important to remark that the complete action,  (\ref{CSB-complete-action-1plus1splitting}), is a bulk action. It contains the information of all the time-like hypersurfaces that foliate the space-time region $\mathcal{M}$, though they are independent among them since we explicitly demand \emph{the condition of the restriction to the boundary} so there are no degrees of freedom propagating from one slice $\mathcal{ B}_{R}$ to the next one $\mathcal{ B}_{R+\delta R}$.

Though a bit extensive, the Hamiltonian analysis of the complete action (\ref{CSB-complete-action-1plus1splitting}) help us to see clearly how the degrees of freedom emerge. And how the symmetries in the bulk are related with the ones in the boundary, so far a family of independent time-like hypersurfaces, that once we chose a particular one, we can call it boundary.

Also note that the main result for the Chern-Simons action does not depend on the particular choice of an time-like hypersurfe (at a particular R). And this come from the fact that we are working with a topological theory at the bulk.


\section{Projecting the equations of motion}\label{Appendix-Projecting-Bulk-EOM}

If, for some reason, we are interested \emph{only} on the dynamics (e.g. equations of motion) on this family of time-like surfaces, it is enough to project the bulk equations of motion and restrict ourselves to these hypersurfaces. 

Our goal in this work is to construct the corresponding theory at the boundary, we focused ourselves in finding an action principle at the boundary, from which we find additional information besides the equations of motion, i.e. the  constraint algebra, the gauge symmetries and degrees of freedom. So we left this discussion for this appendix, where we briefly show how we obtain the same equations of motion as in Sec (\ref{Section-Analyzing the theory at the boundary}), directly by projecting the corresponding bulk ones into the boundary and restricting the dynamics to it. This discussion may shed light on the meaning of the \emph{condition of the restriction to the boundary} introduced in Sec (\ref{Section:Constructing the theory at the boundary}). 

The bulk equations of motion coming from the 3-dimensional Abelian Chern-Simons (\ref{CSAction-components}) are, $\tilde{\epsilon}^{\mu \nu \rho} F_{\nu \rho} =0 $. We can use the same idea for projecting the bulk action but this time for the bulk equations of motions, so we can use the following relation $\tilde{\varepsilon}^{\mu \nu \rho} = 3R^{[\mu}\tilde{\varepsilon}^{ \nu \rho]} \mathrm{d} R$,
\begin{eqnarray}
0= \tilde{\epsilon}^{\mu \nu \rho} F_{\nu \rho} &=& F_{\nu \rho}  3R^{[\mu}\tilde{\varepsilon}^{ \nu \rho]} \mathrm{d} R\\
&=& F_{\nu \rho} \left( R^{\mu}\tilde{\varepsilon}^{ \nu \rho} + R^{\rho}\tilde{\varepsilon}^{ \mu \nu } + R^{\nu}\tilde{\varepsilon}^{ \rho \mu} \right)  \mathrm{d} R\\
&=& \left[ R^{\mu} F_{\nu \rho} \tilde{\varepsilon}^{\nu \rho} - 2 R^{\rho} R_{\rho \nu} \tilde{\varepsilon}^{\mu \nu }  \right]  \mathrm{d} R,
\end{eqnarray}
using that $R^{\rho} F_{\rho \nu} = \partial_{\nu} (R^{\rho} A_{\rho}) - \pounds_{\vec{R}} A_{\nu} $, the previous equation becomes,
\begin{equation}\label{CS-EOM-projection-bulk-appendix}
R^{\mu} F_{\nu \rho} \tilde{\varepsilon}^{\nu \rho} - 2 \left[ \partial_{\nu} \left( R^{\rho} A_{\rho}\right) - \pounds_{\vec{R}} A_{\nu}  \right] \tilde{\varepsilon}^{\mu \nu } =0.
\end{equation}

This equation is still a bulk equation, and completely equivalent to $\tilde{\epsilon}^{\mu \nu \rho} F_{\nu \rho} =0 $. We have only expressed it in terms of their components along the tangential and normal directions. If we want to separate it in two equations corresponding to the tangential and normal parts, we need to contract the previous equation with $A_{\mu}$ and $R_{\mu}$ respectively. Contracting eq. (\ref{CS-EOM-projection-bulk-appendix} ) with $R_{\mu}$,
\begin{equation}
\underbrace{R_{\mu} R^{\mu}}_{\sigma = 1} F_{\nu \rho} \tilde{\varepsilon}^{\nu \rho} - 2 \left[ \partial_{\nu} \left( R^{\rho} A_{\rho} \right)- \pounds_{\vec{R}} A_{\nu}  \right] \underbrace{R_{\mu} \tilde{\varepsilon}^{\mu \nu }}_{=0} =0 \,\,\,\, \Rightarrow F_{\nu \rho} \tilde{\varepsilon}^{\nu \rho} =0,
\end{equation}
and contracting with $A_{\mu}$ and using the previous equation,
\begin{equation}
- 2 A_{ \mu} \left[ \partial_{\nu} \left( R^{\rho} A_{\rho} \right)- \pounds_{\vec{R}} A_{\nu}  \right] \tilde{\varepsilon}^{\mu \nu } =0.
\end{equation}
If we want to restrict the dynamics to the time-like hypersurfaces, and just keep the projected part tangential to the surfaces, we need to impose that,
\begin{equation}
A_{\mu} \pounds_{\vec{R}} A_{\nu}  \tilde{\varepsilon}^{\mu \nu }=0.
\end{equation}
This term encodes the information of the propagation of the degrees of freedom along $\vec{R}$ (the normal direction to the time-like hypersurfaces). By making it zero we are restricting the dynamics to the time-like hypersurfaces without propagation from one time-like hypersurface to the next one.

So the equations of motion describing the dynamics on each of the time-like hypersurfaces are:
\begin{eqnarray}
F_{\nu \rho} \tilde{\varepsilon}^{\nu \rho} &=&0,\\
\label{equation of motion without lambda} A_{ \mu} \partial_{\nu} \left(R^{\rho} A_{\rho}\right)  \tilde{\varepsilon}^{\mu \nu } &=&0,\\
A_{\mu} \pounds_{\vec{R}} A_{\nu}  \tilde{\varepsilon}^{\mu \nu }&=&0. \label{CS-projected-EOM-appendix-ConditionRestrictionBoundary}
\end{eqnarray}
The last equation correspond to the \emph{the condition of the restriction to the boundary} that we identify from the bulk lagrangian to construct the theory at the boundary, that in this context can be seen as the part of the equations of motion containing the information of the propagation in the direction normal to the hypersurfaces. Which it is natural to make zero if we do not want degrees of freedom to propagate in the normal direction (i.e. outside the hypersurfaces).

These equations of motions are equivalent to the ones found in section \ref{Subsection-Well-posedness-action-Boundary-conditions}, and in particular (\ref{equation of motion without lambda}), corresponds to (\ref{CSB-4thEOM}).

\acknowledgments

I'm grateful to Professors Glenn Barnich and Marc Henneaux for their academic and financial support, for encouraging me to develop these ideas, and for enlightening discussions and feedback. To the research group ``Mathematical Physics of Fundamental Interactions'' at the Université Libre de Bruxelles (ULB) for their marvellous hospitality and stimulating environment, to CONACyT (Mexico) and to Solvay Institutes (Belgium) for financial support through a postdoctoral scholarships at ULB (Belgium). I also want to thank ``Benem\'erita Universidad Aut\'onoma de Puebla" for financial support this lasts months through a ``Convenio beca''.




\begin{thebibliography}{99}
	

	
	\bibitem{Ashtekar-2015} A. Ashtekar, \emph{Geometry and physics of null infinity}, \href{https://dx.doi.org/10.4310/SDG.2015.v20.n1.a5}{\emph{Surveys in Differential Geometry} {\bf 20},no. 1 (2015) 99-122.} arXiv:\href{https://arxiv.org/abs/1409.1800}{1409.1800v2 [gr-qc].} 


	\bibitem{Ashtekar-Engle-Sloan-2008} A. Ashtekar, J. Engle and D. Sloan, \emph{Asymptotics and Hamiltonians in a first-order formalism}, \href{https://doi.org/10.1088/0264-9381/25/9/095020
	}{\emph{Class. Quantum Grav.} {\bf 25}, no. 9 (2008) 095020.} arXiv: \href{https://arxiv.org/abs/0802.2527}{0802.2527v2 [gr-qc].} 
	
	\bibitem{Ashtekar-Pretorius-Ramazanoglu-2011} A. Ashtekar, F. Pretorius and F. Ramazanoglu, \emph{Surprises in the evaporation of two dimensional black holes}, \href{https://doi.org/10.1103/PhysRevLett.106.161303}{\emph{Phys. Rev. Lett.} {\bf 106}, 161303 (2011)}. arXiv: \href{https://arxiv.org/abs/1011.6442v1
	}{1011.6442v1 [gr-qc]} 
	\emph{Evaporation of two dimensional black holes}, \href{https://doi.org/10.1103/PhysRevD.83.044040
	}{\emph{Phys. Rev. D} {\bf 83}, (2011) 044040.} arXiv: \href{https://arxiv.org/abs/1012.0077
	}{1012.0077v2 [gr-qc]} 
	
	\bibitem{Ashtekar-Taveras-Varadarajan-2008} A. Ashtekar, V. Taveras and M. Varadarajan, \emph{Information is not lost in the evaporation of 2-dimensional black holes},\href{https://doi.org/10.1103/PhysRevLett.100.211302}{\emph{ Phys. Rev. Lett.} {\bf 100}, (2008) 211302}. arXiv: \href{https://arxiv.org/abs/0801.1811v2}{0801.1811v2 [gr-qc]}. 
	
	\bibitem{Barbero-Diaz-Margalef-Bentabol-Villasenor-2019} J. F.  Barbero, B. D\'iaz, J. Margalef-Bentabol and E. J. S. Villase\~nor, \emph{Dirac's algorithm in the presence of boundaries: a practical guide to a geometric approach}, \href{https://doi.org/10.1088/1361-6382/ab436b}{\emph{Class. Quantum Grav. } {\bf 36}, 20 (2019) 205014.} arXiv: \href{https://arxiv.org/abs/1904.11790}{arXiv:1904.11790v3 [gr-qc]}
	
	\bibitem{Barnich-Troessaert-2010} G. Barnich, and C. Troessaert \emph{Symmetries of asymptotically flat four-dimensional spacetimes at null infinity revisited}, \href{https://doi.org/10.1103/PhysRevLett.105.111103
	}{\emph{Physical review letters} {\bf 105}, no. 11 (2010) 111103.} arXiv:\href{https://arxiv.org/abs/0909.2617v2}{0909.2617v2 [gr-qc].}  
	
	
	\bibitem{Bojowald-2010} M. Bojowald, \href{https://doi.org/10.1017/CBO9780511921759}
	{\emph{Canonical gravity and applications: cosmology, black holes, and quantum gravity}}, Cambridge University Press, 2010.  

	
	\bibitem{Brown-Henneaux-1986} J. D. Brown, and M. Henneaux. \emph{On the Poisson brackets of differentiable generators in classical field theory}, \href{https://doi.org/10.1063/1.527249}{\emph{Journal of mathematical physics} {\bf 27}, no. 2 (1986) 489-491.} 

	
	\bibitem{Buffenoir-Henneaux-Noui-Roche-2004} E. Buffenoir, M. Henneaux, K Noui and P. Roche, \emph{Hamiltonian analysis of Plebanski theory}, \href{https://doi.org/10.1088/0264-9381/21/22/012}{\emph{Class. Quantum Grav.} {\bf 21}, no. 22 (2004) 5203.} arXiv: \href{https://arxiv.org/abs/gr-qc/0404041}{gr-qc/0404041v2.}    
	

	\bibitem{Carrol-2004} Carroll, Sean M. "An introduction to general relativity: spacetime and geometry." Addison Wesley, San Francisco, California, US (2004).

	
	\bibitem{Celada-Gonzalez-Montesinos-2015} M. Celada, D. González, and M. Montesinos, \emph{Alternative derivation of Krasnov’s action for general relativity}, \href{https://doi.org/10.1103/PhysRevD.92.044059}{\emph{Physical Review D} {\bf 92}, no. 4 (2015): 044059.} arXiv: \href{https://arxiv.org/abs/1509.00076}{1509.00076v1 [gr-qc].} 
	

	\bibitem{Compere-2011} G. Compere and F. Dehouck, \emph{Relaxing the parity conditions of asymptotically flat gravity}, \href{https://doi.org/10.1088/0264-9381/28/24/245016}{\emph{Class. Quantum Grav.} {\bf 28}, no. 24 (2011): 245016.} arXiv: \href{https://arxiv.org/abs/1106.4045}{1106.4045v3 [hep-th].}
	

	\bibitem{Corichi-Reyes-2015} A. Corichi and J. D. Reyes, \emph{The gravitational Hamiltonian, first order action, Poincaré charges and surface terms}, \href{https://doi.org/10.1088/0264-9381/32/19/195024}{\emph{Class. Quantum Grav.} {\bf 32} no.19 (2015) 195024.} arXiv: \href{https://arxiv.org/abs/1505.01518}{1505.01518v2 [gr-qc].}  
	
	
	\bibitem{Corichi-Rubalcava-2015} A. Corichi and I. Rubalcava-García, \emph{Energy in first order 2+ 1 gravity}, \href{https://doi.org/10.1103/PhysRevD.92.044040}{\emph{Physical Review D} {\bf 92}, no. 4 (2015): 044040.} arXiv: \href{https://arxiv.org/abs/1503.03030}{1503.03030v2 [gr-qc].} 

	
	\bibitem{Corichi-Rubalcava-Vukasinac-Review} A, Corichi, I. Rubalcava-Garc\'ia and T. Vukasinac, \emph{Actions, topological terms and boundaries in first-order gravity: A review}, \href{https://doi.org/10.1142/S0218271816300111}{Int. J. Mod. Phys. D \textbf{25}, (2016) 1630011}. arXiv:  \href{https://arxiv.org/abs/1604.07764}{1604.07764v1 [gr-qc].}  

	\bibitem{Corichi-Vukasinac-2019} A. Corichi, and T. Vukašinac, \emph{Hamiltonian analysis of a topological theory in the presence of boundaries}, \href{https://doi.org/10.1142/S0218271819500755}{\emph{International Journal of Modern Physics D} {\bf 28} 06 (2019) 1950075.} arXiv: \href{https://arxiv.org/abs/1809.09248v1}{1809.09248v1 [hep-th]} 
	
	\bibitem{Corichi-Vukasinac-2020} A. Corichi and T. Vukasinac, \emph{Canonical analysis of field theories in the presence of boundaries: Maxwell+ Pontryagin}, \href{https://doi.org/10.1088/1361-6382/ab778f}{Classical and Quantum Gravity (2020) Accepted for publication.} arXiv: \href{https://arxiv.org/abs/2001.06068}{2001.06068v1 [gr-qc]} 
	
	\bibitem{Diaz-Higuita-Montesinos-2014} B. D\'iaz, D. Higuita and M. Montesinos, \emph{Lagrangian approach to the physical degree of freedom count}, \href{https://doi.org/10.1063/1.4903183}{\emph{J. Math. Phys.} {\bf 55}, no. 12 (2014) 122901.} arXiv: \href{https://arxiv.org/abs/1406.1156}{1406.1156v2 [hep-th].}    
	
	
	\bibitem{Diaz-Montesinos-2018} B. D\'iaz and M. Montesinos, \emph{Geometric Lagrangian approach to the physical degree of freedom count in field theory}, \href{https://doi.org/10.1063/1.5008740}{\emph{J. Math. Phys.} {\bf 59}, no. 5 (2018) 052901.} arXiv: \href{https://arxiv.org/abs/1710.01371}{1710.01371v2 [gr-qc].}  
	
	
	\bibitem{Dunne-1999} G. V. Dunne, \emph{Aspects of chern-simons theory} in \href{https://doi.org/10.1007/3-540-46637-1_3}{\emph{Topological aspects of low dimensional systems}, Springer, Berlin, Heidelberg, (1999) 177-263.} arXiv: \href{https://arxiv.org/abs/hep-th/9902115}{hep-th/9902115v1}  

	
	\bibitem{Elitzur-etal-1989} S. Elitzur, G. Moore, A. Schwimmer and N. Seiberg,  \emph{Remarks on the canonical quantization of the Chern-Simons-Witten theory}, \href{https://doi.org/10.1016/0550-3213(89)90436-7}{\emph{Nuclear Physics B} {\bf 326}, no. 1 (1989) 108-134.} 
	
	
	\bibitem{Escalante-Carbajal-2011} A. Escalante and L. Carbajal,\emph{ Hamiltonian study for Chern-Simons and Pontryagin theories}, \href{https://doi.org/10.1016/j.aop.2010.09.004}{\emph{Annals of Physics} {\bf 326}, no. 2 (2011) 323-339.} arXiv: \href{https://arxiv.org/abs/1107.4023}{1107.4023v1 [math-ph].} 
	
	
	\bibitem{Gallardo-Montesinos-2011} A. Gallardo and M. Montesinos, \emph{The boundary field theory induced by the Chern–Simons theory}, \href{https://doi.org/10.1088/1751-8113/44/13/135402}{\emph{Journal of Physics A: Mathematical and Theoretical} {\bf 44}, no. 13 (2011) 135402.} arXiv: \href{https://arxiv.org/abs/1008.4883}{1008.4883v2 [hep-th].}   

	
	\bibitem{Geiller-2017} M. Geiller, \emph{Edge modes and corner ambiguities in 3d Chern–Simons theory and gravity}, \href{https://doi.org/10.1016/j.nuclphysb.2017.09.010}{\emph{Nuclear Physics B} {\bf 924} (2017) 312-365.} arXiv: \href{https://arxiv.org/abs/1703.04748}{1703.04748v5 [gr-qc].} 
	

	\bibitem{Gourgoulhon-2012} E. Gourgoulhon, \emph{3+1 Formalism in General Relativity: Bases of Numerical Relativity}, \href{https://doi.org/10.1007/978-3-642-24525-1}{Lecture Notes in Physics \textbf{846}, Springer (2012).} 
	

	\bibitem{Grumiller-Merbis-Riegler-2017} D. Grumiller, W. Merbis, and M. Riegler, \emph{Most general flat space boundary conditions in three-dimensional Einstein gravity}, \href{https://doi.org/10.1088/1361-6382/aa8004}{\emph{Class. Quantum Grav.} {\bf 34}, no. 18 (2017) 184001.} arXiv: \href{https://arxiv.org/abs/1704.07419}{1704.07419v2 [hep-th].} 

	
	\bibitem{Henneaux-Troessaert-2018electro} M. Henneaux, and C. Troessaert, \emph{Asymptotic symmetries of electromagnetism at spatial infinity}, \href{https://doi.org/10.1007/JHEP05(2018)137}{\emph{J. High Energ. Phys.} {\bf 2018} (2018) 137.} arXiv: \href{https://arxiv.org/abs/1803.10194}{1803.10194v2 [hep-th].} 
	
	
	\bibitem{Henneaux-Troessaert-2018ADM} M. Henneaux and C. Troessaert, \emph{BMS group at spatial infinity: the Hamiltonian (ADM) approach}, \href{https://doi.org/10.1007/JHEP03(2018)147}{\emph{J. High Energ. Phys.} {\bf 2018}, no. 3 (2018) 147.} arXiv: \href{https://arxiv.org/abs/1801.03718}{1801.03718v1 [gr-qc].}  
	
	
	\bibitem{Henneaux-Teitelboim-1994} M. Henneaux and C. Teitelboim, \emph{Quantization of Gauge Systems}, Princeton University Press, Princeton, New Jersey, (1994). 

	\bibitem{Krasnov-2011} K. Krasnov, \emph{Pure connection action principle for general relativity}, \href{https://doi.org/10.1103/PhysRevLett.106.251103}{\emph{Physical review letters} {\bf 106}, no. 25 (2011): 251103}. arXiv: \href{https://arxiv.org/abs/1103.4498v1}{1103.4498v1 [gr-qc]} 
	
	\bibitem{Maldacena1999} J. Maldacena, \emph{The large-N limit of superconformal field theories and supergravity}, \href{https://doi.org/10.1023/A:1026654312961}{\emph{International journal of theoretical physics} {\bf 38}, no. 4 (1999) 1113-1133.} arXiv: \href{https://arxiv.org/abs/hep-th/9711200}{hep-th/9711200v3.} 
	
	
	\bibitem{Papantonopoulos-2011} E. Papantonopoulos, ed. \emph{From gravity to thermal gauge theories: The AdS/CFT correspondence}. \href{https://doi.org/10.1007/978-3-642-04864-7}{\emph{Lecture Notes in Physics} {\bf 828.} Springer Science and Business Media, (2011).} 
	
	
	\bibitem{Peldan-1994} P. Peldan,\emph{ Actions for gravity, with generalizations: a title}, \href{https://doi.org/10.1088/0264-9381/11/5/003}{\emph{Class. Quantum Grav.} {\bf 11}, no. 5 (1994) 1087.} arXiv: \href{https://arxiv.org/abs/gr-qc/9305011}{gr-qc/9305011v1.}
	
	
	
	\bibitem{Poisson-2004} E. Poisson, \emph{A relativist's toolkit: the mathematics of black-hole mechanics}, \href{https://www.cambridge.org/mx/academic/subjects/physics/cosmology-relativity-and-gravitation/relativists-toolkit-mathematics-black-hole-mechanics?format=HB&isbn=9780521830911}{Cambridge university press, 2004.}  
	
	
	\bibitem{Romano-1993} J. D. Romano, \emph{Geometrodynamics vs. connection dynamics}, \href{https://doi.org/10.1007/BF00758384}{\emph{General relativity and gravitation} {\bf 25}, no. 8 (1993) 759-854.} arXiv:\href{https://arxiv.org/abs/gr-qc/9303032}{gr-qc/9303032v1.}
	
	
	\bibitem{Romero-Vergara-2002} J. M. Romero and J. D. Vergara, "Boundary conditions as constraints." arXiv: \href{https://arxiv.org/abs/hep-th/0212035}{hep-th/0212035v2} 
	
	
	\bibitem{Rubalcava-Juarez-2019a} I. Rubalcava-Garcia and J. Juarez-Susano, "Hamiltonian analysis and boundary terms for local and non-local theories". In preparation 
	

	\bibitem{Sheikh-Jabbari-Shirzad-2001} M. M. Sheikh-Jabbari and A. Shirzad, \emph{Boundary conditions as Dirac constraints}, \href{https://doi.org/10.1007/s100520100590}{\emph{The European Physical Journal C - Particles and Fields} {\bf 19} 2 (2001) 383-390.} arXiv: \href{https://arxiv.org/abs/hep-th/9907055v3}{hep-th/9907055v3.} 
	
	
	\bibitem{Temple-Lysov-Mitra-Strominger-2015} H. Temple, V. Lysov, P. Mitra, and A. Strominger, \href{https://doi.org/10.1007/JHEP05(2015)151}{\emph{BMS supertranslations and Weinberg’s soft graviton theorem}, \emph{Journal of High Energy Physics} {\bf 2015} 5 (2015) 151.} arxiv: \href{https://arxiv.org/abs/1401.7026}{1401.7026v2 [hep-th].} 
	

	\bibitem{Tong-2016} D. Tong, Lectures on the Quantum Hall Effect, arXiv:\href{https://arxiv.org/abs/1606.06687}{1606.06687v2 [hep-th]} 
	
	
	\bibitem{GFTC-2011-Book} G.F. Torres del Castillo, \emph{Differentiable manifolds: a theoretical physics approach}, \href{https://doi.org/10.1007/978-0-8176-8271-2}{Birkh\"auser, Springer Science, 2011.} 
	
	
	\bibitem{Troessaert-2013} C. Troessaert, \emph{Canonical structure of field theories with boundaries and applications to gauge theories}, (2013). arXiv: \href{https://arxiv.org/abs/1312.6427}{1312.6427v2 [hep-th].} 
	
	
	\bibitem{Zanelli-2008-UsesCS} J. Zanelli, \emph{Uses of Chern‐Simons Actions}, \href{https://doi.org/10.1063/1.2971999}{\emph{ AIP Conference Proceedings}, {\bf 1031} 1 (2008) 115-129.} Arxiv: \href{https://arxiv.org/abs/0805.1778}{0805.1778v2 [hep-th].} 
	
	
	\bibitem{Zanelli-2012-LecturesCS} J. Zanelli, \emph{Introductory lectures on Chern-Simons theories}, \href{https://doi.org/10.1063/1.3678608}{\emph{AIP Conference Proceedings}, {\bf 1420} 1, (2012 ) 11-23.} 
	
	
	\bibitem{Zanelli-2012-CSinGravitation} J. Zanelli, \emph{Chern–Simons forms in gravitation theories}, \href{https://doi.org/10.1088/0264-9381/29/13/133001}{\emph{Class. Quantum Grav.} {\bf 29}, 13 (2012) 133001.} arXiv: \href{https://arxiv.org/abs/1208.3353}{1208.3353v1 [hep-th].}
	
	
	\bibitem{Witten-1988a} E. Witten, \emph{Topological gravity}, \href{https://doi.org/10.1016/0370-2693(88)90704-6
	}{\emph{Physics Letters B} {\bf 206} 4 (1988) 601-606.} 


	\bibitem{Witten-1988b} E. Witten, \emph{2+ 1 dimensional gravity as an exactly soluble system}, \href{https://doi.org/10.1016/0550-3213(88)90143-5
	}{\emph{Nuclear Physics B} {\bf 311} 1 (1988) 46-78.}
	
	
	
\end{thebibliography}
\end{document}